\pdfoutput=1
\RequirePackage{ifpdf}
\ifpdf 
\documentclass[pdftex]{sigma}
\else
\documentclass{sigma}
\fi

\usepackage{mathrsfs,amsbsy}

\newcommand{\const}{{\mathrm{const}}}

\newcommand{\hpsi}{\hat{\Psi}}
\newcommand{\hchi}{\hat{\chi}}

\newcommand{\dv}{\mathrm{d}^{3}}

\def\mmm{\mathfrak{m}}
\def\nnn{\mathfrak{n}}

\def\xx{{\mathrm{x}}}

\def\1{1 \!\! 1}

\newcommand{\DF}{{\mathcal{D}}}

\newcommand{\id}{\mathbb{I}}
\newcommand{\vev}{{\it vev~}}
\newcommand{\Aa}{{\mathcal{A}}}

\numberwithin{equation}{section}

\begin{document}

\allowdisplaybreaks

\renewcommand{\thefootnote}{$\star$}

\renewcommand{\PaperNumber}{027}

\FirstPageHeading

\ShortArticleName{Emergent Models for Gravity: an Overview of Microscopic Models}

\ArticleName{Emergent Models for Gravity:\\ an Overview of Microscopic Models\footnote{This
paper is a contribution to the Special Issue ``Loop Quantum Gravity and Cosmology''. The full collection is available at \href{http://www.emis.de/journals/SIGMA/LQGC.html}{http://www.emis.de/journals/SIGMA/LQGC.html}}}

\Author{Lorenzo SINDONI}

\AuthorNameForHeading{L.~Sindoni}

\Address{Max Planck Institute for Gravitational Physics, Albert Einstein Institute, \\
 Am M\"uhlenberg 1, 14467 Golm, Germany}
\Email{\href{mailto:sindoni@aei.mpg.de}{sindoni@aei.mpg.de}}

\ArticleDates{Received October 02, 2011, in f\/inal form May 04, 2012; Published online May 12, 2012} 

\Abstract{We give a critical overview of various attempts to describe gravity as an emergent phenomenon, starting from examples of condensed matter physics, to arrive to more sophisticated pregeometric models. The common line of thought is to view the graviton as a~composite particle/collective mode. However, we will describe many dif\/ferent ways in which this idea is realized in practice.}


\Classification{83C45; 83C50; 81Q20}

\renewcommand{\thefootnote}{\arabic{footnote}}
\setcounter{footnote}{0}

\tableofcontents

\section{Introduction}

Despite some compelling theories have been proposed to f\/inally merge quantum theory with gravitation \cite{Polchinski, RovelliBook}, it is fair to say that we are still away from a def\/initive answer to the question ``what is the microscopic structure of spacetime?''.

Interestingly enough, classical general relativity does possess in itself some peculiar signatures that link it to thermodynamics: from the most celebrated results about black hole thermodyna\-mics~\cite{Bardeen, Bekenstein} to the more recent results establishing a connection between thermodynamics of local horizons and general relativity (as well as other theories of gravity) \cite{Chirco:2010,Chirco:2009,ElingGuedensJacobson,Jacobson,Padmanabhan:2003,Padmanabhan:2009,Verlinde}, the AdS/CFT correspondence~\cite{Aharony} and the gravity/f\/luid connection (see, e.g.,~\cite{Hubeny}).
These results show that it is not unconceivable that general relativity is ultimately a form of hydrodynamics for the microscopic constituents whose collective, large scale behavior is described by geometrodynamics.

The objective of this review paper is to present a number of approaches in which gravity-like features emerge (in dif\/ferent degrees of ref\/inement and complexity) in various models, in specif\/ic regimes. Rather than following a single line of thought, focussing on a single kind of models, or a historical perspective,
we will review a certain number of results trying to follow the logic f\/low that will lead us from the most simple to the more complicated, questioning at each stage the gains and the penalties to be payed. This is done in order to show the intricate web of relations among the most dif\/ferent approaches, trying to establish relations among them and to make easier the assessment of their pros and cons, but also disentangling the deep structural issues from the accidental realizations that are due to the specif\/ic property of the model.

While the area of emergent gravity could be considered as a subject per se, it can be considered a part of the research in quantum gravity, given that it might provide precious insights for the problem of obtaining the correct semiclassical limit in a quantum gravity theory. In that context, we will have to describe how the microscopic degrees of freedom are organized by their dynamics, highlighting, when possible, the general properties of the state or of the regime leading, at large distances, to a smooth Lorentzian manifold with quantum matter f\/ields living over it.

For this purpose, the various techniques developed in the realm of (classical and quantum) statistical mechanics, many body problem in the various incarnations encountered in condensed matter physics, the renormalization group analysis and all the related ideas (recently, motivated also from the systems relevant for quantum information)
will be (most likely) decisive in the elaboration of concrete strategies to solve the issue. This has already been shown in the concrete example of dynamical triangulations for Euclidean quantum gravity, of their Lorentzian counterpart, causal dynamical triangulations \cite{CDT2,CDT1,CDT4,CDT3, CDT0}, and of matrix models for two dimensional gravity coupled to certain classes of conformal f\/ields \cite{MM2, MM1}.

We will not try to enter too much in the philosophical aspects of the idea of emergence, given that the very
def\/inition is unclear. As a provisional def\/inition, we will intend as emergence of a given theory as
a reorganization of the degrees of freedom of a certain underlying model in a~way that leads to a
regime in which the relevant degrees of freedom are qualitatively dif\/ferent from the microscopic
ones.

In absence of a unambiguous notion of emergence upon which a universal agreement can be
reached, we will avoid carefully to discuss its general principles, leaving them very vague, and instead we will move directly to the
discussion of the concrete models, proposed in the past years to explain the long
range phenomenon called gravity. We will focus on the merits of each proposal, as well as the shortcomings. For dif\/ferent perspectives on the idea of emergent gravity, the reader can see, for instance, \cite{Dreyer2,Dreyer1,BLH,HuMicro,HuHydro,Hu:2011,OritiDepth, Seiberg}.

{\bf Disclaimer.} The nature of this review is very specif\/ic. It is focused on the methods, ideas and results of various models in which gravitational dynamics is recovered, more or less successfully, from a non-gravitational system.
As a consequence, we will not provide a review of many important results that refer to the thermodynamical behavior of gravity, that we have already mentioned. These important results pertain to
the thermodynamical layer of the description. We will rather focus the attention layer immediately below, i.e.\ on the way the microscopic degrees of freedom get assembled to produce gravity, whose interest for quantum gravity is obvious. For a discussion of the thermodynamical aspects of gravity, we refer to the extant literature.

This said, we will leave out of the analysis one relatively important topic on the emergence of spacetime, the AdS/CFT scenario mentioned earlier. For this very important holographic perspective we will refer, in addition to the references already given, to
\cite{Balasubramanian,Berenstein,Koch,Gherghetta,Horowitz, Lin}, where various dif\/ferent aspects
are discussed, with special emphasis on the origin of gravity from CFTs.

The plan of the paper is as follows. We will start in Section~\ref{appetizer} by recalling simple ideas of composite gauge bosons from fermion/antifermion pairs, and some key ideas of induced gravity. In Section~\ref{analogue} we will brief\/ly recall the main results of analogue models, which have provided valuable inspiration for this whole approach. In Section~\ref{Lorentz} we will consider carefully the case for Lorentz invariance, highlighting some important dif\/f\/iculties that are plaguing analogue models with many components. In Section~\ref{BEC} we will give an account on how Bose--Einstein condensates (BEC) can be used to have dynamical analogues of Newtonian gravity. While practically of little interest, this allows a f\/irst concrete (actually literal) example of gravity as hydrodynamics and of the emergence of the equivalence principle. We will pass then to the description of some ideas related as the point of view that the graviton is a Goldstone boson associated to the spontaneous breaking of Lorentz invariance, in Section~\ref{goldstoneboson}, while in Section~\ref{diffeo} we will examine the possibilities that dif\/feomorphism, gravitons and their dynamics might emerge from a dif\/ferent sort of phase transitions appearing in some lattice systems. We will also summarize the key ideas that are behind the notion of ``emergent gauge invariance''. In Section~\ref{pregeometry} we will then describe some models in which there is no background geometry, where the metric genuinely emerges as a composite f\/ield, while in Section~\ref{signature} we will provide some examples in which the signature of spacetime is determined dynamically.
Concluding remarks and future perspectives are presented in Section~\ref{challenges}.

We will use the signature $(-,+,+,+)$ unless otherwise specif\/ied, and we will use units $\hbar=1$, $c=1$, apart from the sections involving BECs, for which the usual nonrelativistic sets of units are used.

\section{A warm up}\label{appetizer}

The question about the fundamental nature of the gravitational f\/ield, geometry and the spacetime
manifold is just the most extreme case of the question about the fundamental nature of the
interactions (and of matter f\/ields) that we observe. It is therefore interesting to start by considering two very simple cases in which gauge symmetry and gravitational dynamics can emerge. These two examples will represent two prototypes on which many proposals have been based. Since they also allow us to discuss a number of important points, we start our tour from them.

\subsection{Bjorken's model for electrodynamics}

The idea that the fundamental interactions can be, after all, not so fundamental, is not a new idea in theoretical physics. In a rather old work, Bjorken \cite{Bjorken:1963vg,Bjorken:2001pe}  has shown how it might be possible that a $U(1)$ gauge interaction like electromagnetism can be generated in a purely fermionic theory with a four fermions interaction. The idea is to start from Nambu--Jona Lasinio-like fermionic systems \cite{NJL1,NJL2} described by the following generating functional\footnote{Incidentally, NJL models were designed to reproduce some features of the BCS theory of superconducti\-vity~\cite{BCS}, in particular the formation of fermion-antifermion pairs, to be applied in the study of high energy physics.}
\begin{gather*}
Z[J_{\mu}] = \int \mathcal{D}\psi \mathcal{D}\bar{\psi} \exp\left(-\frac{i}{\hbar}\left(S[\psi,\bar{\psi}]-\int \bar{\psi} \gamma^{\mu}\psi J_{\mu} d^4 x\right)\right) ,
\end{gather*}
where the action is given by
\begin{gather*}
S[\psi,\bar{\psi}] = \int d^{4}x \left(\bar{\psi}  i\gamma^{\mu} \partial_{\mu} \psi - m \bar{\psi}\psi+\frac{G}{2} (\bar{\psi}\gamma^{\nu}\psi)^{2} \right),
\end{gather*}
with $G$ being the coupling constant associated to the four fermions vertex. Notice that this generating functional allows us to compute correlation functions for the composite operator
\begin{gather*}
A^{\mu}=\bar{\psi} \gamma^{\mu}\psi,
\end{gather*} which transforms as a Lorentz vector.

Of course, this model is non-renormalizable by power counting, and hence it must be seen as an ef\/fective theory valid at suf\/f\/iciently small energies (with respect to the energy scale set by~$G$).
A suitably def\/ined Hubbard--Stratonovich transformation
leads to the rewriting of this in terms of the composite f\/ield alone
\begin{gather*}
Z[J_{\mu}] = N' \int \mathcal{D}A \exp\left(-i \int d^{4}x \left( \frac{A^{2}}{2G} - V(A-J) \right) \right),
\end{gather*}
where $V$ is the ef\/fective potential, and it is given, formally, by
\begin{gather*}
i \int d^{4}x V(A) = \log \det(i \gamma^{\mu} (\partial_{\mu} +A_{\mu})-m) -\log \det(i \gamma^{\mu} \partial_{\mu} -m).
\end{gather*}
By means of this integration over closed fermionic loops, the f\/ield $A_{\mu}$ inherits its own dynamics. The lowest order ef\/fective action generated in this way gives
\begin{gather*}
S_{\rm ef\/f} = \int d^{4}x \left( \frac{A^{2}}{2G} + \frac{1}{4e^{2}} F_{\mu\nu}F^{\mu\nu} +\frac{\lambda}{4}\big(A^{2}\big)^{2}\right),
\end{gather*}
where $e$ is a coupling constant given by
\begin{gather*}
\frac{1}{e^{2}}= \frac{1}{12 \pi^{2}} \log \frac{\Lambda^{2}}{m^{2}},
\end{gather*}
with $\Lambda$ an UV cutof\/f regulating the integrals (ultimately related to the energy at which
the ef\/fective theory breaks down).

The resulting ef\/fective theory for the composite vector f\/ield is not gauge invariant, since there are terms containing $A^{2}$ in the ef\/fective action. Nevertheless, the potential might lead to non-vanishing~\vev of~$A_{\mu}$. In this case, the theory would give spontaneous Lorentz symmetry breaking. The massless Nambu--Goldstone modes associated to this symmetry breaking might be interpreted as photons, having the same kind of coupling to the fermions, i.e.\ minimal coupling.
For a detailed discussion, see also \cite{Birula}.

Such an example shows how it is possible to have composite vector mediators arising from four-fermions interactions (see also \cite{Banks:1980rh, Eguchi,EguchiSugawara,Terazawa1,Terazawa2}). In the past, models applying this kind of reasoning to the case of the emergent linearized graviton have been put forward \cite{Atkatz, Ohanian}, implementing the idea that the linearized graviton might arise as a certain composite f\/ield of a nongravitational f\/ield theory. However, the models have dif\/f\/iculties in going beyond the linear order to include self coupling of the emergent gravitons, thus barring the way to regain general relativity even in a perturbative sense (see, for instance \cite{Deser} and references therein). Furthermore, the question whether a four Fermi theory can be truly fundamental, and not only an ef\/fective f\/ield theory valid below a certain scale, requires further clarif\/ications\footnote{There are mainly two problems. First, power counting argument suggest that the theory is perturbatively nonrenormalizable. Second, these theories might saturate and then violate the unitarity bound for scattering amplitudes. Of course, there might be a natural solution to these issues without invoking gauge invariance, within the framework of asymptotically safe theories. However, the matter is not yet settled.}.

Lorentz symmetry breaking is unavoidable, in this class of models. From a pure group-theoretic point of view, the photon is described by a helicity one state. In absence of gauge invariance,
a mass term for a vector f\/ield cannot be forbidden. Therefore, one needs another mechanism to keep the vector massless. Viewing the photon as a Nambu--Goldstone mode is certainly an
option. However, the price to pay is that it has to come from a vector theory, and hence it has to
break Lorentz invariance. Given the tight constraints on Lorentz invariance violation (LIV), this option seems to require a lot of work on the fundamental model to be really viable (essentially, a viable mechanism to suppress the LIV operators in the ef\/fective f\/ield theory). We will come back to this point later.

There are other two key dif\/ferences from a genuine gauge theory. First of all, besides the massless
Nambu--Goldstone modes, there are other polarizations for such composite vector f\/ields. They
are massive, with mass determined by the shape of the potential $V(A)$. If the potential is steep enough in these components, they are not observable at low energy.

Second, the fact that a theory is gauge invariant implies Schwinger--Dyson and Ward--Takahashi identities among the correlation functions, and, consequently, specif\/ic patterns in the renormalization of the theory.
These are the quantum version of the classical equations of motion and of the conservation equations following from the presence of symmetries. A proposal for the emergence of gauge invariance has to reproduce these identities, at least within certain approximations that determine the regime of validity of the model itself.

Despite all these shortcomings, Bjorken's model is a prototype for all the f\/ield-theoretic models (i.e.\ described by some form of f\/ield theory on some notion of background spacetime manifold) of emergent gauge systems as we know them.
In this model, the gauge boson is interpreted as a vacuum expectation value of a composite operator of certain fundamental f\/ields, whose dynamics generate a strong-coupling phase which
is most ef\/fectively described by these composite f\/ields. In a sense, the resulting equations of motion (classical or quantum) might be viewed as Landau--Ginzburg equation for the order parameter of the specif\/ic phase of the theory.

\subsection{Induced gravity versus emergent gravity}

Is it possible to generalize the mechanism to gravity? Before considering this very general point, it is interesting to mention the mechanism of induced gravity.
In the previous example, a central role is played by quantum corrections to the tree level action. It is instructive to see how these can be relevant within a context in which gravity is introduced.

In \cite{Sakharov}, Sakharov suggested the idea that the dynamics of the gravitational f\/ield could be seen as a manifestation of the ``elasticity of spacetime'' with respect to the matter f\/ields living over it. For a technical presentation and discussion of the various subtleties involved, see \cite{VisserSakharov}. Assume to have a certain set of matter f\/ields, denoted collectively by $\phi$. Assume that they propagate over a metric $g_{\mu\nu}$ which is, at least at a f\/irst glance, just an external f\/ield, without any sort of dynamics. The only equations of motion relevant, at the tree level, are
\begin{gather*}
\big(\Box_{g}-m^{2} \big) \phi =0,
\end{gather*}
for the matter f\/ields\footnote{Of course, for fermions one should consider the corresponding Dirac equation and consistently take into account the small dif\/ferences with respect to the bosonic case. Small modif\/ications are present for particles with dif\/ferent spins. Nevertheless, the general result is unchanged.}. However, quantum corrections signif\/icantly change the picture.

By simple manipulations of the functional integrals, it can be seen that loop corrections due to the dynamics of the matter f\/ields can generate an ef\/fective action for the gravitational f\/ield.
At one loop:
\begin{gather*}
S_{1-{\rm loop}}[g] \approx \log \det \left( \Box_{g} - m^{2} \right).
\end{gather*}
The standard Schwinger--DeWitt technique \cite{Buchbinder} shows that this term involves a number of divergent contributions, which need appropriate counterterms to be renormalized. In particular, among these terms there are what we call the cosmological constant term and the Einstein--Hilbert action.

This picture has been further ref\/ined \cite{Adler:1980pg,Adler:1982ri, Zee1,Zee2}, connecting the problem of emerging gravitational dynamics with the breaking of scale invariance and the appearance of energy scales separating UV from IR degrees of freedom.
In the same spirit, but with a slightly dif\/ferent twist, there are proposals of some models of gravity (and gauge interactions) entirely arising from fermionic/bosonic loop corrections.
These attempts are certainly very important in the discussion of the nature of gravitational interaction, and must be taken into account. We will devote an entire section to them later in this paper.

Induced gravity is certainly an option that should be considered. Nevertheless, it is perhaps not appropriate to think of induced gravity as an issue of emergent gravity. Indeed, there is a dif\/ference between induced gravity and models {\it \'a la} Bjorken: the mechanism described by Bjorken is essentially the formation of a composite f\/ield, by some sort of fermion/antifermion pairs condensation:
\begin{gather*}
A_{\mu} \propto \langle \bar{\psi} \gamma_{\mu} \psi \rangle.
\end{gather*}
Subsequently, the dynamics of the f\/ield $A_{\mu}$ is induced by the quantum corrections.
The case of induced gravity is fundamentally dif\/ferent in the f\/irst step: the pseudo-Riemannian structure of~spacetime is postulated {\it a priori}. It is assumed that the matter f\/ields are living on some sort of spacetime. What is not assumed is the fact that the metric tensor does obey some equation of motion. Sakharov's induced gravity, then, is a case in which only the gravitational dynamics is induced by quantum corrections, and not the entire geometrical picture. Stated in another way, in Sakharov's picture, the graviton is not necessarily a composite particle (even if it is implicitly assumed so).

In this sense, then, it is important to distinguish this scheme from a more radical emergent picture, where all the concepts needed to speak about spacetime geometry are emerging from a~pre-geometric phase.

\section{Emergent spacetimes: lessons from condensed matter}\label{analogue}

Despite the remarkable observational successes of GR, the gravitational f\/ield, due to its relative weakness, remains very hard to be tested in regimes of strong curvature (where the behavior of gravity signif\/icantly deviates from the Newtonian description and requires a fully relativistic description) and simultaneously under experimentally controllable situations, where, at least in principle, the separation between contamination ef\/fects and the physical ef\/fect is clear.

Besides the obvious importance of the strong curvature regime for the dynamics of the gravi\-tational
f\/ield itself, there are a number of kinematical ef\/fects that are relevant for the dynamics of the matter f\/ields living in the spacetime manifold. There is a f\/irst obvious question that arises, in curved spacetimes, and it is the way in which matter f\/ields feel the underlying geometry. A~feature such as nonminimal coupling should be more visible in strong curvature regimes (even though it might have surprising impact in certain low curvature but large size systems~\cite{NMC}).

Besides the issue of the validity of the equivalence principle there are some enormously important phenomena, such cosmological particle creation and Hawking radiation, that
have a~huge impact on our  understanding of the origin of structures in inf\/lationary scenarios, and on the fate of unitary evolution in quantum mechanics when black holes can form
\cite{Mathur}.

Despite the fact that to date there are dif\/ferent ways to approach the theoretical aspects of these ef\/fects, still, their status as proper phenomena is unclear. A direct detection is needed to conf\/irm the theoretical calculations. However, gravitational f\/ields that are strong enough to lead to particle creation phenomena are not available in a laboratory. With``true" gravitational f\/ields, the exploration of strong curvature seems to be limited to the mere observation of astrophysical or cosmological phenomena, subject to all the related uncertainties and limitations.

However, it is possible to circumvent these dif\/f\/iculties, at least partially, and work with ``artif\/icial'' curved spacetime metrics even in a not particularly sophisticated laboratory. Indeed, it has been
realized by Unruh  \cite{Unruh:1980cg} that the pattern of propagation of waves in a moving f\/luid strongly resembles
the propagation of waves in curved spacetimes.
The general idea behind this fact is that, when a f\/luid is f\/lowing with a given velocity f\/ield, sound waves are dragged by the f\/luid f\/low. Hence, if a supersonic f\/low is realized in some region, sound waves cannot go upstream. This is in analogy with what happens with trapped regions in general relativity. Using this very simple idea it can be easily realized that some sort of sonic analogue of black holes spacetimes can be realized.

This general expectation can be formalized in a rigorous mathematical theorem \cite{Visser93}, which shows how in some situations the properties of sound propagation can be exactly described by a curved pseudo-Riemannian metric.

{\bf Theorem.} Consider an inviscid, barotropic and irrotational f\/luid. Sound waves, i.e.\ li\-nea\-rized perturbations around a~given f\/luid conf\/iguration, are solutions of a Klein--Gordon equation for a massless scalar f\/ield living on a curved metric,
\begin{gather*}
g_{\mu\nu} =  \frac{\rho}{c_s}   \begin{pmatrix} -(c_s^2-v^2) &\vdots& -\vec{v} \\
\cdots& \cdot& \cdots \\
-\vec{v} &\vdots &\mathbb{I}_3 \end{pmatrix}.
\end{gather*}
This metric tensor is called the \textit{acoustic metric}. It depends algebraically on the properties of the f\/luid f\/low: the local density $\rho$, the velocity f\/ield $\vec{v}$, and the speed of sound $c_s$.
Therefore, the dynamics for this metric is not described by some sort of Einstein equations written in terms of some curvature tensors. Rather, the acoustic metric is a solution of the f\/luid equations, i.e.\ the continuity equation and Euler equation.

During the last years this result has been extended to other condensed matter systems, like non-linear electrodynamic systems, gravity waves, etc..
For the present discussion it is important to keep in mind that there are many condensed matter physics, from rather
ordinary ones like nonrelativistic f\/luids to more exotic ones like quantum f\/luids and graphene layers that can exhibit a peculiar phenomenon: in certain conditions, to be specif\/ied according to the nature of the system itself, the excitations, the perturbations or other signals do propagate, at least at low energies/frequencies, over ef\/fective Lorentzian geometries (or with ef\/fective Dirac operators, if fermions are involved), even if the original condensed matter system is relativistic under the Galilean relativistic group of transformations.  This result has been variously revisited, extended, applied to many dif\/ferent systems with very dif\/ferent properties, and it has been developed into a whole branch of research, investigating the properties and the possibilities of the so called
``analogue models for gravity''. Since this is not a review on analogue models, we refer to the recent review \cite{AnalogueReview} which presents an up-to-date picture of the investigation.
For a detailed discussion of $^{3}\mathrm{He}$ models (and for a thorough discussion of the Fermi point scenario), see \cite{VolovikDroplet} where some implications for emergent gravity scenarios are also discussed.

The bottom line is that one could imagine to realize {\it effective curved spacetimes} in systems which could be studied experimentally (and not just observationally) in a laboratory, thus opening the possibility of detecting and testing directly some features of QFT in curved spacetimes, like Hawking radiation and cosmological particle creation.

A crucial point must be kept in mind: these models are thought to be used to study ef\/fects that involve just the fact that quantum f\/ields live on curved spacetimes, the latter not being necessarily solutions of some sort of Einstein equations. Indeed, these geometries are determined by the equations of motion of the underlying condensed matter system, which have structural dif\/ferences with respect to normal dif\/feomorphism invariant systems of PDE that make a one to one correspondence not possible, in general.

It should be said that analogue models are not just an exercise in mathematical physics.
In fact, analogue models can provide relevant insights and useful ideas about those situations where spacetime is emergent from a complicated underlying quantum gravity dynamics.
In the next chapter it will be shown that the ideas we have already discussed in the area of analogue models have a counterpart in many theoretical frameworks designed to describe the semiclassical limit of quantum spacetime near the Planck scale, as well as their phenomenological consequences.

\subsection{BEC systems}
A particularly appealing system for the purposes of analogue gravity is represented by Bose--Einstein condensates. A BEC is a particular phase of a system of identical bosons in which a~single energy level has a macroscopic occupation number. They are realized by using ultra-cold atoms (see \cite{blackbook}).
Under these extreme conditions, quantum phenomena become a key ingredient in determining the macroscopic properties of the system.
The analysis of these systems, therefore, opens a new area of investigation for quantum f\/ields in curved spacetime \cite{BarceloBEC, Garay1,Garay2}.

We do not want to enter a full f\/ledged analysis of the theoretical understanding of the physics of BECs. for which we refer to the extant literature. Rather, we recall a number of basic facts that serve as a guide for the understanding of the physical content of the
models and their relevance. We do so, since many features will be common to other approaches.
Also, within the landscape of emergent gravity models, the BEC is perhaps the model in which the
transition between the microscopic domain and the long wavelength limit, displaying emergent gravitational phenomena, is the cleanest.

The simplest description of a BEC system (specif\/ically, for dilute Bose gases) is the mean f\/ield description provided by the condensate wavefunction, a complex scalar (but can (and has to) be generalized to higher spins) classical function encoding the f\/luid properties of the condensate.
The equations of motion for the many body problem can be translated into an ef\/fective equation for this function, the so-called Gross--Pitaevski equation, which is essentially a nonlinear Schr\"odinger equation
\begin{gather*}
i\hbar\frac{\partial \psi}{\partial t} = -\frac{\hbar^{2}}{2m} \nabla^{2} \psi - \mu \psi - \kappa |\psi|^{2} \psi,
\end{gather*}
where $m$ is the mass of the bosons, $\mu$ the chemical potential, $\kappa$ controls the strength of the two body interactions (the dominant scattering channel, in the dilute gas approximation), and $\psi$ is the condensate wavefunction.
By means of the Madelung representation $\psi = \sqrt{n}\exp(-i\theta)$, this complex equation can be split into two real equations, the continuity equation for the number density $n$ and the Bernoulli equation for the velocity potential $\theta$.
With respect to the hydrodynamical equations for a perfect f\/luid, there is a correction term, known as the quantum potential or the quantum pressure, that generates an additional contribution to the pressure of the f\/luid. It is entirely quantum in nature, as signaled by the presence of $\hbar$ in its expression
\begin{gather*}
V_{q} = -\frac{\hbar^{2}}{2m}\frac{\nabla^2n^{1/2}}{n^{1/2}}.
\end{gather*}
A key role is played by the coherence length of the system, known as the healing length
\begin{gather*}
\xi = \frac{\hbar}{(2m\mu)^{1/2}}.
\end{gather*}
It plays many roles: it determines the typical dynamical scale of the condensate, as well as the properties of the dispersion relation of the phonons. This latter, known as the Bogoliubov dispersion relation, has the form:
\begin{gather}
\omega^{2}(k) = c_s^2 k^2 + \frac{\hbar^2k^4}{4m^2}.
\label{BogoliubovDR}
\end{gather}
The healing scale controls the transition between the relativistic low energy dispersion relation $\omega \approx c_{s} k$ to the high energy behavior $\omega \propto k^{2}$.

In the low energy regime, where the quantum potential term is essentially inactive, we can show that the phonons do
propagate in ef\/fective curved acoustic Lorentzian geometries. Furthermore, the fact that in the case of BECs quantum properties are directly relevant for the physics of phonons explains their relevance in simulating quantum phenomena in curved (acoustic) spacetimes. Indeed, the low energy ef\/fective f\/ield theory of the phonons is a quantum theory of a scalar f\/ield on the acoustic spacetime def\/ined by the condensate. Despite being a very simple case, it is enough to reproduce, at least in principle, interesting phenomena of particle creation in curved spacetime.  For a recent and extensive discussion of BEC models, see \cite{Jannes}.

The most important part in designing an analogue model is to control the physical properties of the system itself in such a way to reproduce the desired acoustic metric. In the case of the perfect f\/luid we have discussed above, the procedure is pretty straightforward: by controlling the velocity prof\/ile one can obtain the various properties of the acoustic metric. In the case of the condensate, an alternative way has been envisaged to achieve this goal.
In the case of the BEC the scattering length, related to the interatomic interactions, can be modif\/ied by acting directly on the condensate, by means of the so-called Feshbach resonances \cite{blackbook}. By using this technique, one can make the speed of sound a position dependent function. Therefore, instead of keeping f\/ixed the speed of sound and changing the velocity prof\/ile, in order to have subsonic and supersonic regions it is enough to keep the f\/low's velocity f\/ixed and to change the speed of sound.

The possibility of simulating black hole spacetimes and hence to have some sort of grasp on the Hawking ef\/fect has been put forward since the works \cite{Garay1,Garay2} (see also \cite{BarceloBEC}).
Similarly, it has been proposed that BEC could be used to realize analogue of expanding Friedman--Robertson--Walker (FRW) spacetimes, and hence to have the possibility of check in a laboratory phenomena which are typical of inf\/lationary scenarios, like particle creation \cite{BLVFRW, bosenovae}.
In this case, it becomes manifest how important is the possibility of controlling the scattering length in realizing a curved acoustic metric.

The method of Feshbach resonance is so ef\/fective that one could realize physical situations of rather exotic phenomena which do not have an analogue in the standard approach to quantum f\/ield theories, namely, signature changes events \cite{sign-chg}. By tuning the value of the scattering length, and in particular making it negative, one can immediately realize that $c_{s}^{2}<0$, which means that the acoustic metric becomes a Riemannian one (signature $(++++)$).

BEC-based analogue models, therefore, do represent a very nice opportunity to discuss many dif\/ferent phenomena which are of interest for theoretical physics.
There are two fundamental gains from the study of these systems. First of all, testing phenomena of quantum f\/ield theo\-ries in curved spacetimes which have been only predicted and never measured directly, given the dif\/f\/iculty of detecting them in strong gravitational f\/ields generated by the astrophysical sources, is an important achievement \textit{per se}, making theoretical prediction subject to experimental verif\/ication ({see also \cite{Jannes} for a dedicated overview}).

An additional reason of interest is related to the so called trans-Planckian problem. In the derivation of Hawking radiation and the spectrum of inf\/lationary perturbations, one is implicitly assuming that the ef\/fective f\/ield theory description (QFT in curved spacetime) is holding up to arbitrarily high energies/arbitrarily small scales (for a discussion and references see, for instance~\cite{transplanckian}, but see~\cite{Mathur} for a dif\/ferent perspective).
This is certainly an untenable assumption. Almost certainly, at suf\/f\/iciently small scales, the EFT picture will break down, higher dimensional operators will become more and more relevant up to the point that the fundamental theory of spacetime (quantum gravity/strings/\dots) might be required to perform the calculations.

The trans-Planckian problem, then, could open the door to this (mostly unknown) physical ef\/fects on low energy predictions. In this respect, it is important to understand how much the predictions of Hawking's spectrum and particle creation in an inf\/lationary model are af\/fected by modif\/ications to the UV behavior of the theory. In the case of BEC, we are in complete control of the trans-Planckian physics of the system, i.e.\ the properties of the model below the healing length.

As it has been shown, phonons do not have an exact relativistic dispersion relation: the dispersion relation receives corrections which are more relevant the higher is the energy, with the healing length scale playing the role of the separation scale between the relativistic branch and the trans-phononic branch of the spectrum. The robustness of Hawking radiation and particle creation against high energy modif\/ied dispersion relations has been discussed in several works \cite{Barbado,Barcelo1,Barcelo2,Brout:1995wp,Corley:1996ar,Finazzi1,Unruh:1994je,Unruh:2004zk}. Similarly, while a signature change event would lead to an inf\/inite particle production, the presence of high energy modif\/ications of the spectrum regularizes the divergencies, predicting a f\/inite result~\cite{sign-chg}.

The potentialities of these analogue systems are further highlighted by some works where the issue of the backreation of the emitted radiation on spacetime  (i.e.\ the precise way in which Hawking radiation leads dynamically to the evaporation of the black hole) is explicitly consi\-de\-red~\cite{Balbinotbackreaction2, Balbinotbackreaction1}.

Recently, rather important progresses have been done on both the experimental side and on the numerical side. In fact, very recently it has been discussed the realization of a BEC analogue of a black hole~\cite{BHBEC}. Even though this is certainly encouraging, still this is not enough for the purposes of detecting directly Hawking quanta. On the other hands, the production of Hawking radiation from an acoustic black hole in a BEC has been detected in numerical simula\-tions~\cite{Carusotto}, while experiments with other analogue models have shown signals of analogue Hawking radiation~\cite{Faccio,Weinfurtner}.
Finally, some numerical estimates on the ef\/fects of cosmological particle creation have been done in~\cite{Silke-Infla} (see also~\cite{Prain}).

It is clear that the underlying physics, for these analogue models, is non-relativistic: some Lorentz-violating ef\/fects are gradually turned on as the energy is increased. Even though this is a possibility, this might not be the case for ``real'' spacetime, and it might be that other kind of UV completions of our low energy ef\/fective f\/ield theories would lead to unexpectedly large ef\/fects on the low energy predictions of Hawking spectrum and cosmological particle production. To decide this matter, of course, a complete calculation within a theory of quantum gravity is needed.

\subsection{Fermi point scenario}
It is interesting to brief\/ly mention here some important work done on fermionic systems. It is hard to overestimate the importance of the discussion of analogue systems containing fermionic degrees of freedom, given that these are among the basic ingredients of the Standard Model. As a second instance, it gives intriguing hints that there might be a deep connection between the particle spectrum, their interactions and topology, broadening then the perspective of emergent systems.

The analysis of the excitations of fermionic system over the vacuum state involves the determination of the properties of the Fermi surface (even in a generalized sense) in momentum space. Indeed, it has been shown that the properties of the excitations are essentially tied to the topological properties of the Fermi surface \cite{Horava2005,Volovik2001, VolovikDroplet} (see also \cite{Volovik2008, Volovik2007} for more recent presentations). We refer to these references for more detailed accounts of the idea.

As a consequence, for fermionic systems the determination of the dispersion relations, and hence of the emergent low energy spacetime symmetry, amounts to the determination of certain topological properties of the Fermi surface (in the generalized sense of the region of the momentum space where the ef\/fective propagator has poles).
This classif\/ication is robust, in the sense that only topological reasonings are needed, and many details of the underlying condensed matter system are simply not needed. This is of course a form of universality.

In particular, it has been established that the relevant universality class is the one associated with the presence of Fermi points, in momentum space (we refer to the literature for all the important details).
In this case, the inverse ef\/fective propagator, for an excitation of momentum~$p_{\mu}$ reads
\begin{gather*}
G^{-1}(p) = e^{\mu}_{a} \Gamma^{a}\big(p_{\mu}-p_{\mu}^{0}\big) + \text{corrections},
\end{gather*}
where $p_{\mu}^{0}$ is the position of the Fermi point in momentum space $\Gamma^{a}$ are just the four Hermitian $2\times 2$ matrices $(\id_2,\sigma^i)$ and $e^{\mu}_{a}$ plays the role of
a tetrad. In such a conf\/iguration, then, for momenta close enough to the Fermi point, the energy spectrum is linear, and essentially one recovers a notion of Lorentz invariance. The correction terms in the inverse propagator are responsible for the restoration of Galilean invariance at high energy, but they can be neglected, at low energy, as one can neglect the corrections to the relativistic behavior in the Bogoliubov dispersion relation for the phonons in a BEC.

In fact, one can push a little bit more the analysis and show that it is conceivable to expect that the matrix $e^{\mu}_{a}$ and the vector $p_{\mu}^{0}$ become position dependent. In this case one has the generalization of the result obtained for f\/luids and phonons in BEC: the excitation do propagate in a nontrivial metric-af\/f\/ine background for momenta close to the Fermi point. Indeed, the tetrad gives rise to a position-dependent metric tensor
\begin{gather*}
g^{\mu\nu} = e^{\mu}_{a}e^{\nu}_{b} \eta^{ab},
\end{gather*}
with the connection being just a $U(1)$ connection
\begin{gather*}
A_{\mu}(x) \sim p_{\mu}^{0}(x),
\end{gather*}
which might represent a form of electromagnetic f\/ield.
The entire construction can be generalized suitably in order to obtain the emergence of nonabelian gauge f\/ields by choosing the correct topology (i.e.\ universality class) of the quantum vacuum.
This picture gives a strong suggestion on the fact that the entire set of gauge interactions (gravity included) might be low energy remnants of an underlying microscopic theory, with the low energy theory, its symmetries and its dynamical content partially determined by topological properties of the quantum vacuum.

However, as in the case of BEC and, partially, of Sakharov's idea, this is only half of the problem. Indeed, this tetrad and these gauge f\/ields appear as nondynamical classical background f\/ields: they are related to the collective f\/ields that would describe the vacuum state, and they would be determined by the same quantum equations of motions that determine the vacuum. Hence, if we
start from a nonrelativistic system containing nonrelativistic fermions, the equations for these geometrical and gauge backgrounds will not be of the form of an Einstein--Yang--Mills system.

Furthermore, as we will discuss brief\/ly in the next section, in the case in which several species are present there is no guarantee that at low energy there will be just one notion of ef\/fective metric (see the discussion in Section~6.4.3 of \cite{Volovik2001} for this specif\/ic point). We refer again to the papers and to the book mentioned above for additional references and extensive discussions on the potential relevance of these ideas for an emergent theory of gravity, especially on the role of the various energy scales involved.

\section{The fate of Lorentz invariance}\label{Lorentz}

The examples discussed above suggests that Lorentz invariance\footnote{We will often refer to the Lorentz group even though the Poincar\'e group should be mentioned instead. We apologize for this abuse of terminology.} that we observe and with which we build the Standard Model, might be, after all, only a symmetry property associated to the low energy/large distances regime that we probe with our experiments or observations. Instead the fabric of spacetime at smaller and smaller distances could be characterized by a dif\/ferent symmetry group, or by a totally dif\/ferent geometrical structure. One immediate manifestation of this would be the phenomenon of dispersion for light rays, i.e.\ the dif\/ference in the speed of propagation of light rays with dif\/ferent frequencies. The possibility that quantum gravity ef\/fects might lead to nontrivial propagation of waves has now a relatively long story \cite{ Alfaro:1999wd, GambiniPullin}, and has f\/lourished in what is called Quantum gravity phenomenology \cite{GACQGP1,GACQGP2,Hossenfelder, MaccioneLiberati}.

Within the context of analogue models, the general procedure to derive the existence of low energy Lorentz invariance involves the manipulations of the f\/ield equations (classical or quantum) for the system under consideration, the imposition of the appropriate boundary conditions and, typically, a restriction of the analysis to a specif\/ic kinematical regime in which one or more conditions are precisely met. This is because, while it is certainly \emph{possible} to have Lorentz invariance emerging in condensed matter systems, the result is not obvious at all.

\subsection{Multi-component systems: normal modes analysis}
So far we have presented results for which it is implicitly assumed that there is just a single f\/ield describing the excitations. This may suf\/f\/ice to investigate phenomena like particle creation, as we have discussed previously, given that the ef\/fect that is under consideration does not really involve the kind of matter f\/ields that are involved. However, it is important to understand how generic is the emergence of a single pseudo-Riemannian metric when several components are present. In addition to the obvious interest in understanding the conditions necessary to have the emergence of a Lorentzian structure, it turns out that this analysis gives indications about possible alternatives. It is instructive to discuss two cases, where the physical features are clear. These will be paradigmatic for the cases in which more f\/ields are involved.

The f\/irst case to be discussed is the propagation of light in a crystal. On scales larger than the size of the cell, a crystal is an homogeneous medium: all its points are equivalent, having the same physical properties. However, not all the directions are equivalent: in general a crystal is anisotropic, the anisotropy being related to the preferred directions selected by the fundamental cell.

Therefore, the propagation of signals, like sound waves, or light, is af\/fected by the symmetry properties. In the case of the propagation of light (for the propagation of sound waves see for instance~\cite{Landau7}), the Maxwell equations in vacuum are replaced by ef\/fective equations where
 (see~\cite{Landau8} for a complete treatment) the overall electromagnetic properties of the material are included in the f\/ields $\mathbf{D}$, $\mathbf{H}$,  related to the electric and magnetic f\/ields $\mathbf{E}$, $\mathbf{B}$ by means of constitutive relations, encoding the microphysics of the system in macroscopic observables.

To make a long story short, in the case of linear constitutive relations with negligible magnetic properties ($\mu_{ij}=\delta_{ij}$), the dispersion relation is provided by the so called Fresnel equation:
\begin{gather*}
\det(n^2\delta_{ik}-n_i n_j-\varepsilon_{ij}) =0,
\end{gather*}
where $\varepsilon_{ij}$ encodes the linear constitutive relation relating the electric f\/ields.
This equation is an equation for the refractive index vector $\mathbf{n}=\mathbf{k}/\omega$. The shape of the space of solutions of this equation depends on the shape of the tensor $\varepsilon$, which encodes the EM properties of the medium we are considering.
For instance, if $\varepsilon_{ij}=\varepsilon \delta_{ij}$ we have an isotropic medium, and we can see that~$\mathbf{n}$ lives on the sphere of radius $\varepsilon^{1/2}$. The geometric interpretation is obviously given in term of a single Riemannian metric, leading to a single pseudo-Riemannian structure for the propagation of light signals, with the speed of light in vacuum replaced by the speed of light in the crystal.

In the case of uniaxial crystals, we have that $\varepsilon_{ij} = \mathrm{diag}(\varepsilon_1,\varepsilon_1,\varepsilon_2)$. This leads to a Fresnel equation which is factorizable:
\begin{gather*}
\big( n_x^2 + n_y^2 +
        n_z^2-\varepsilon_1 \big) \big(  \varepsilon_{1} n_x^2 + \varepsilon_1 n_y^2 + \varepsilon_2 n_z^2 - \varepsilon_1 \varepsilon_2\big)=0.
\end{gather*}
Therefore, this equation gives rise to two bilinear forms, with which one can def\/ine two pseudo-Riemannian metrics, thus leading to a bi-metric theory. The two photon polarizations are traveling at dif\/ferent speeds. This phenomenon is called bi-refringence.

In the case of biaxial crystals, we have three principal axis with three distinct eigenvalues, the factorization of the Fresnel determinant into two quadratic terms is no longer possible: we have to solve an algebraic equation of the fourth degree.
In the three dimensional vector space where the vector $\mathbf{n}$ lives, the surface def\/ined by this equation is a complicated self-intersecting quartic surface. The features of this surface are the origin of the interesting optical properties of this class of crystals~\cite{BornWolfe}.

Notice the relation between the symmetry group of the crystal and the corresponding geometrical structure: the more anisotropic is the crystal, the more we break Lorentz invariance in the analogue model. In the most general case, Lorentz invariance is not an approximate symmetry in these analogue systems, even at low energies.

Another interesting class of analogue systems with a rather rich phenomenology is given by the two components BEC systems~\cite{Liberati:2005id,Weinfurtner:2006wt}. These are just a special case of the so-called {\it spinor condensates} \cite{spinorcondensate}. In these systems, there are two species of bosons, rather than only one as in the BEC discussed previously. Similar considerations also apply to the Fermi point scenario, as we have mentioned in the previous section.

These two components can be thought to be two dif\/ferent hyperf\/ine levels of the same atom. The two components are interacting, typically by laser coupling, which is inducing transitions between the two levels. The analysis of the spectrum  of the quasi-particles is considerably more involved than the single component case. Nevertheless, in the hydrodynamic regime (when the quantum potential is neglected), the spectrum can be obtained analytically. For the detailed calculation, which is based on a suitable generalization of the Madelung representation, see~\cite{Liberati:2005id}.

\looseness=1
The Hamiltonian describing the 2BEC contains a number of parameters: the masses of the atoms, the chemical potentials, the strengths of the atomic interaction, and the coupling describing the induced transitions between the components. According to the values of these parameters, the phononic branch of the spectrum shows dif\/ferent behaviors. In particular, one can distinguish three geometrical phases:
for generic values of the parameters, the analogue geometry is not Riemannian, but rather Finslerian (see next subsection). If some tuning is made, it is possible that, instead of this Finslerian structure, a bi-metric (pseudo-Riemannian) will describe the propagation of the phonons (which are of two kinds, in these 2BEC systems). Finally, if more tuning is performed by the experimenter on the various constants describing the properties of the system, the two families of phonons perceive the same Lorentzian geometry at low energy. In this specif\/ic corner of the parameter space Lorentz invariance as an approximate symmetry is recovered at low energies.

Of course, this clean description holds only in the hydrodynamic limit, when the quantum potentials are neglected. As in the case of single BEC, the higher the energy, the large will be the contribution of the corrections to the dispersion relations (see \eqref{BogoliubovDR}). Again, as we shall prove later, dispersion can be seen as the manifestation of Finsler geometry. Therefore, in 2BECs there are two ways in which we can see Finsler geometry emerge: f\/irst, by a generic choice of the physical properties of the system, and secondly by the appearance of dispersion.

\subsection{From analogue models to Finsler geometry}

Analogue models have shown that it is certainly conceivable that Minkowski spacetime or, more generically, Lorentzian geometries might arise in some ef\/fective regime of an otherwise nonrelativistic system (in the sense of special and general relativity). This feature, however, is not generic. If we take a general system with many components, extending the analysis of the physical examples that we have considered, the normal modes analysis \cite{Barcelo:2001cp} shows that each mode will have, in general, dif\/ferent speed of propagation. In the typical case, having $n$ modes, the dispersion relation can be expressed as
\begin{gather*}
 \omega^{2n} = Q_{i_1\dots i_2n}k^{i_1}\cdots k^{i_2n}.
\end{gather*}
In \looseness=1 general this equation def\/ines a (possibly self-intersecting) algebraic hypersurface of deg\-ree~$n$ in momentum space. Given the homogeneity of the relation under rescalings $k\rightarrow \lambda k$, \mbox{$\omega\rightarrow \lambda \omega$}, this hypersurface is nothing else than the generalization of the null cone of special relati\-vity.

The microscopic details of the system are encoded in the tensor $Q$. According to this, there are dif\/ferent emergent geometrical structures. Tuning the microscopic parameters, it might happen that the tensor $Q$ factorizes into a product of metric tensors:
\begin{gather*}
 Q_{i_1\dots i_2n} = g^{(1)}_{i_1i_2}\cdots g^{(n)}_{i_{n-1}i_{n}},
\end{gather*}
In this case, the emergent spacetime structure is pretty clear: there is an ef\/fective multimetric structure, where each mode is propagating on a dif\/ferent light cone determined by the corresponding metric structure. By additional tuning, it might be possible to reduce this multi-metric structure to a single Lorentzian structure, when $g^{(1)}=g^{(2)}=\cdots =g^{(n)}$. We can say, then, that the emergence of an ef\/fective Lorentzian spacetime is not generic: some conditions have to be satisf\/ied in order for a single metric to be recovered. This is often translated in terms of symmetries of the underlying model. For instance, anisotropies in crystals are directly related to the appearance of birefringence.

Of course, even in the most symmetric case, the emergent Lorentz invariance is only an approximate symmetry of the spectrum. For instance, in condensed matter systems the underlying spacetime symmetry is Galilean invariance, which has rather dif\/ferent properties than Lorentz invariance. This fact is often contained into corrections to the lowest order/low energy equations for the perturbations. In systems like BEC, the quantum potential gets more important the larger is the energy of the phonon, modifying the linear dispersion relation in the way described by the Bogoliubov dispersion relation \eqref{BogoliubovDR}. Dispersion is in general a feature that we should expect in analogue spacetimes. Of course, dispersion is not compatible with a notion of pseudo-Riemannian geometry.

However, the discussion made so far should have clarif\/ied that in analogue models the most general situation is the one in which the tensor $Q$ is not factorized. In this case, while the dispersion relation is still homogeneous (hence there is no dispersion), it is generally anisotropic: modes moving in dif\/ferent directions will propagate with dif\/ferent speeds. Again, this feature cannot be described by Lorentzian geometry.

\looseness=-1
It is interesting that, despite that dispersion and anisotropic propagation do not have a~direct interpretation in terms of Lorentzian geometry, they do have a geometrical interpretation in terms of a suitable metric generalization of Riemannian geometry, namely Finsler geo\-met\-ry~\cite{baochernshen, Barcelo:2001cp,rund}.

\looseness=1
This is not the place to present a review on Finsler spaces. A rather exhaustive presentation of the most relevant technical points, with a list of references, is given in~\cite{tesi}.
For the moment it will suf\/f\/ice to say that a Finsler space is a manifold endowed with a \mbox{(pseudo-)norm} in the tangent space of each point of the manifold.
Clearly, Riemannian manifolds are special cases of Finsler spaces, being those Finsler manifolds where the norm is derived from a~scalar product. They can be also characterized in terms of symmetries: maximally symmetric Finsler spaces (i.e.\ those with the maximal number of independent Killing vectors) are indeed maximally symmetric Riemannian spaces. In other words, Finsler spaces are in general locally anisotropic metric spaces, while the locally isotropic ones are the ordinary Riemannian spaces~\cite{AdlerBazin,Laugwitz}.

Therefore, the appearance of Lorentzian or genuinely Finslerian structures in the ef\/fective descriptions of emergent spacetime ultimately depends on the degree of isotropy of space, or other, more general, symmetry requirements. Notice also that even in a spatially isotropic situation with dispersion, the ef\/fective spacetime geometry is Finslerian~\cite{GLS}.

Research on the concrete impact of Finsler geometry is an ongoing ef\/fort. The interested reader can see \cite{Hehl:2005,Punzi1,Punzi2,Schuller,SkakalaTesi,Skakala3,Skakala1,Skakala2,Skakala4,Vacarubook} for some concrete proposal to def\/ine Finslerian structures in the Lorentzian signature to open the road for a concrete investigation of the potential implications on phenomenology.

From a purely geometrical standpoint, Finsler spaces are perfectly reasonable geometrical backgrounds to be used for the def\/inition of distances and other metrical properties, as well as for the propagation of particles and f\/ields. However, from the physical point of view, anisotropic Finsler spaces suf\/fer from the problems of Lorentz violating theories in reproducing the observed phenomenology.

The idea that spacetime structure at the Planck scale might deviate considerably from the f\/lat Minkowski background of special relativity is recurrent theme in models of quantum gravity. Despite no def\/inite proof exist that Lorentz symmetry breaks down in a quantum gravity model (see however the already mentioned results \cite{ Alfaro:1999wd, GambiniPullin}), it is conceivable that near the Planck scale spacetime starts deviating signif\/icantly from the continuum dif\/ferential picture of Riemannian geometry to become something dif\/ferent\footnote{Of course, this is only one possibility, despite being the common one in traditional quantum gravity approaches.}. Therefore, it is plausible that the Lorentz group loses its preeminent role as well, to be deformed or replaced by other symmetry groups (see, for instance, \cite{Girelli:2010b, Girelli:2004,Girelli:2010} and references therein).

\looseness=-1
One of the hypothesis that has been considered is that, due to the peculiar microscopic dynamics of spacetime, dispersion relations (or, equivalently, wave equations) might receive Lorentz-breaking corrections. The naive expectation is that, provided that the corresponding operators are Planck-suppressed, Lorentz breaking ef\/fects might be invisible at low energy, in which the theory  f\/lows towards a Lorentz invariant f\/ield theory. However, it has been shown that this is not the case when quantum mechanics is included, in general. Indeed, Lorentz violating operators at the Planck scale are able to contaminate low dimension operators (those relevant at low energies) through radiative corrections \cite{Collins,Serone}, leading to a naturalness problem (essentially, the large hierarchy between the LIV scale and the scales involved in the physical processes that we know so far, or other f\/ine tuning problems) when this phenomenon is compared to the experimental evidences, which put extremely strong constraints on Lorentz violation. Despite some proposals, including supersymmetry \cite{Groot,Mattingly}, the problem is still essentially open.

In absence of strong symmetry arguments (alternative to the direct imposition of Lorentz invariance), multimetric geometrical phases or Finsler geometries are bound to emerge. A~way around this would be to get noncommutative spacetimes, as it has been suggested in the past. However, this would amount to a much more radical departure from the usual setting of dif\/ferential geometry.
The f\/irst lesson that we have to learn, when trying to emerge general relativity from something else, is that even the kinematical setting, i.e.\ Lorentzian metrics, are not automatically arising at low energies, and that a certain amount of work has to be dedicated to the neutralization of all the possible sources of Lorentz symmetry violation\footnote{See, on this issue, the Fermi point scenario discussed by Volovik \cite{VolovikDroplet}. See also \cite{BarceloJannes} for a discussion of the possibility that internal observers might not detect LIV, after all.}.
We will come back to this point in Section~\ref{goldstoneboson}.

Summarizing, in an emergent scenario, the mere emergence of the kinematical background for gravity, i.e.\ Lorentz invariance, is not a trivial business at all, and might lead to an immediate dismissal of the model from the very beginning.

\section{Analogue nonrelativistic dynamics: the BEC strikes back}\label{BEC}

After having brief\/ly exposed the main dif\/f\/iculties of getting the right kinematical framework for gravity (i.e.\ having a properly def\/ined low energy form of Lorentzian metrics) we can move to the next level of dif\/f\/iculty: the emergence of some reasonable
dynamics for the emergent gravitational f\/ield.

Despite the fact that analogue models are designed to reproduce and study phenomena in curved spacetime for which the exact form of the dynamics for the background geometry is immaterial, this does not imply that no analogue dynamics at all can be reproduced.

As an example, it is possible to prove that in the presence of particular symmetries, the equations of f\/luid dynamics can be mapped into the symmetry reduced equations for the metric tensor. This has f\/irst been realized in \cite{Cadoni}, where the case of spherical black holes has been considered. There, it has
been proved that the gravitational f\/ield equations do have a precise correspondence with the constrained steady f\/luid f\/low with an appropriate source.

Furthermore, it is easily realized that a mechanism like Sakharov's induced gravity might be at work in these models.
Here, the loop corrections due to the quantum phononic f\/ield might generate cosmological constant and Einstein--Hilbert terms. Nevertheless, there is a key dif\/ference with the original idea of induced gravity: while in Sakharov's proposal the dynamics of the gravitational f\/ield is produced entirely by means of loop corrections, in a BEC there is already a dynamics for the condensate inducing a dynamics for the acoustic metric, i.e.\ the Gross--Pitaevski equation. In these systems, then, the loop contributions are just subdominant corrections \cite{Barcelo:2001, Visser:2001}.
In this perspective, there have been proposals to use curved space techniques to assess the ef\/fects of backreaction of quantum f\/luctuations on the mean f\/ield description of a condensate \cite{Balbinotbackreaction2, Balbinotbackreaction1}.

BEC analogues do of\/fer another intriguing possibility, that is to genuinely simulate nonrelativistic gravitational theories, like Newtonian gravity \cite{Girelli:2008gc}, and to discuss some structural issues which are quite independent from the specif\/ic form of the gravitational
theory at hand, like the emergence of an equivalence principle \cite{NBEC}.

\subsection{Single BEC}
In the following we will be very specif\/ic regarding the meaning of the analogue gravitational dynamics. We will restrict the investigation to the equivalent of the nonrelativistic limit of general relativity, i.e.\ the weak f\/ield, small speeds limit of Einstein's theory. In this limit the dominant ef\/fect onto the motion of particles is due to the Newtonian potential,
\begin{gather*}
2\Phi_{N} = g_{00} + c^{2},
\end{gather*}
for massive particles. In the case of BEC (and other) analogues, this regime amounts to the case in which the deviations from homogeneous static f\/low are small ($\rho\approx \const$, $\vec{v} \approx 0$).

The obvious dif\/f\/iculty to properly def\/ine a nonrelativistic model, the masslessness of the phonons, can be circumvented by making them massive. This can be done very easily. Masslessness of the phonons in a BEC is a consequence of Goldstone's theorem\footnote{For a detailed account of the concepts involved, the consequences and some applications, see \cite{Burgess}.}: the Hamiltonian of a BEC is invariant under a global $U(1)$ transformation of the second quantized f\/ield operators
creating and destroying particles, which is the symmetry related to the conservation of the total number of bosons.

Breaking this symmetry, making the phonons pseudo-Nambu--Goldstone bosons allow for a~gap in the dispersion relation, i.e.\ a mass term. Of course this implies that a nonconservation of the number of bosons has to be included. The origin of this might be
an interaction with a reservoir of bosons, if we think about atoms. However, one might also imagine that the bosons which are condensing are quasiparticles or other collective excitations that might not be associated to number conservation laws (magnons, spinons, etc.). In this cases, one has to include in the equations of motion additional $U(1)$ breaking terms.

In order to do so, the Hamiltonian $\hat{H}_0$ used for the number conserving system needs to be slightly modif\/ied,
by introducing a term which is (softly) breaking the $U(1)$ symmetry, for instance
\begin{gather*}
\hat H_0 \rightarrow  \hat H = \hat H_0+ \hat H_\lambda, \qquad \hat H_\lambda = - \frac{\lambda}{2} \int \dv\xx \big( \hpsi(\xx) \hpsi(\xx) + \hpsi^{\dagger}(\xx) \hpsi^{\dagger}(\xx) \big).
 \end{gather*}
The parameter $\lambda$  has the same dimension as~$\mu$. This is the most simple addition, relevant for a~dilute gas and for
a system in which the number of bosons is conserved in average.

With this new Hamiltonian, the non-linear equation governing the f\/ield operators becomes
\begin{gather*}
i\hbar \frac{\partial }{\partial t}\hpsi = [\hat H , \hpsi] =  -\frac{\hbar^2}{2m} \nabla^{2}\hpsi - \mu \hpsi  + \kappa |\hpsi|^{2}\hpsi- \lambda \hpsi^{\dagger}.
\end{gather*}

These additional terms are responsible for the appearance of a mass term in the dispersion relation of the phonon, which enables us to def\/ine properly a nonrelativistic limit.
In particular, the dispersion relation over a homogeneous condensate gives
\begin{gather*}
E(p) = \left( \frac{p^{4}}{4 m^{2}} + c_{s}^{2} p^{2} + \mathcal{M}^{2}c_{s}^{4} \right)^{1/2}
\end{gather*}
where
\begin{gather*}
\mathcal{M}^{2} = 4 \frac{\lambda(\mu+\lambda)}{\mu+2\lambda} m^{2}
\end{gather*}
gives the ``rest mass''\footnote{Notice the abuse of language: this term would just give the energy gap at zero momentum. We are using the relativistic language in light of the analogy that has been established.}, and
\begin{gather*}
c_{s}^{2} = \frac{\mu+2\lambda}{1+\lambda} \frac{\kappa n_{c}}{m}
\end{gather*}
with
\begin{gather*}
n_{c} = \frac{\mu+\lambda}{\kappa}
\end{gather*}
the number density of the background condensate.

The second extension that has to be made involves the equations of motion. To properly take into account the ef\/fect of quasiparticles moving over the condensate on the condensate itself, the Gross--Pitaevski equation has to be replaced with
the Bogoliubov--De Gennes equation for the condensate wavefunction. This is obtained by using an improved mean f\/ield approximation
which includes the f\/irst ef\/fects of the noncondensed fraction. This is obtained by writing the second quantized operator in the following approximate form
\begin{gather*}
\hpsi(\xx) \approx \psi(x) \mathbb{I} + \hchi(\xx).
\end{gather*}
The resulting equation for the condensate wavefunction reads
\begin{gather}
i\hbar \frac{\partial \psi}{\partial t} = - \frac{\hbar^{2}}{2m} \nabla^{2} \psi - \mu \psi - \lambda \psi^{*} + \kappa |\psi|^{2}\psi +
2 \kappa\nnn(x) \psi + \kappa\mmm(x) \psi^{*},\nonumber\\
\nnn(x)= \langle \hchi^{\dagger}(x) \hchi(x) \rangle,  \qquad \mmm(x) = \langle \hchi^{2}(x) \rangle,
\label{eq:anomalous}
\end{gather}
where we are neglecting $\langle \hchi^{\dagger} \hchi \hchi \rangle$ which is suppressed by the diluteness of the condensate and the hypothesis of a small noncondensed fraction.

Notice that the form of this equation resembles structurally the semiclassical Einstein equations. However, the phonons are backreacting on the condensate not with their energy but with their number density.

The f\/inal step is to consider the weak f\/ield limit, by restricting the condensate wavefunction to the following regime
\begin{gather}
\psi = \left(\frac{\mu + \lambda}{\kappa} \right)^{2} (1+u(x)),
\end{gather}
where $u(x) \ll1$ is a real function encoding the deviation from perfect homogeneity. One could have introduced a small imaginary component $(1+u(x)) \rightarrow (1+u(x)+iv(x))$. However, the coupling of $v(x)$ to the quasiparticles is weaker than the one of $u(x)$. Furthermore, $u(x)$ has direct meaning in terms of the analogue Newtonian potential
\begin{gather*}
\Phi_{N}(\xx) = -\frac{1}{2} (g_{00}+1) = -\frac{3}{2}c_{s}^{2} u(\xx),
\end{gather*}
while the function $v(\xx)$ is related, through gradients, to the shift vectors of the acoustic metric, when written in the ADM form.

\subsection{Lessons: Poisson equation, mass and the cosmological constant}

In the static limit, when we can neglect temporal gradients, the Bogoliubov--De Gennes equation (which is essentially the equation governing the hydrodynamics of the condensate) can be rewritten in the following form:
\begin{gather}
\left(\nabla^{2}-\frac{1}{L^{2}} \right) \Phi_{N}(\xx) = 4 \pi G_{N} \rho_{\rm matter} +\Lambda,
\label{eq:poisson}
\end{gather}
where we have def\/ined the ``matter'' mass density as
\begin{gather*}
\rho_{\rm matter} = \mathcal{M} \left( \tilde{\nnn} + \frac{1}{2} \mathrm{Re} (\tilde{ \mmm}) \right), \qquad \tilde{\nnn} = \nnn-\nnn_{\Omega}, \qquad \tilde{\mmm} = \mmm-\mmm_{\Omega},
\end{gather*}
with the $\nnn_{\Omega}$, $\mmm_{\Omega}$ being the expectation values of the operators def\/ined in \eqref{eq:anomalous} in the state containing no phonons (i.e.\ with $u(x)=0$), a vacuum term
\begin{gather*}
\Lambda = \frac{2\kappa(\mu+4 \lambda)(\mu+2\lambda)}{\hbar^{2}\lambda} \left( {\nnn}_{\Omega} + \frac{1}{2} \mathrm{Re} ({ \mmm}_{\Omega}) \right),
\end{gather*}
an analogue Newton's constant
\begin{gather*}
G_{N} =  \frac{\kappa(\mu+4\lambda)(\mu+2\lambda)^{2}}{4 \pi \hbar^{2} m \lambda^{3/2} (\mu+\lambda)^{1/2}},
\end{gather*}
and the length
\begin{gather*}
L^{2} = \frac{\hbar^{2}}{4m(\mu+\lambda)},
\end{gather*}
which corresponds to the healing length for this kind of condensate.

Therefore, with this simple models we can simulate a nonrelativistic short range version of the gravitational interaction, obeying a modif\/ied Poisson equation. Unfortunately the range of the interaction is controlled by the healing length, which is a UV scale, in this model. This signals
that the analogy is not good at all.
However, this should not come as unexpected: indeed, especially in ef\/fective quantum f\/ield theories, the mass of the bosonic f\/ields (and hence the range of the interaction they mediate) gets renormalized by contributions which are quadratic and quartic in the cutof\/f of the theory, and hence pushed to the scale of the cutof\/f itself\footnote{Fermions would not suf\/fer from this, provided that some additional custodial symmetry (e.g.\ chiral symmetry) is active.}. At the level of this semiclassical analysis adopted here this is not obviously at work, but the principle is essentially the same: in absence of a custodial symmetry, there is no reason why should we expect a long range/massless analogue gravitational f\/ield.

Of course, this is taken care of by dif\/feomorphism invariance, in general relativity. However, we will see that in a multi-BEC case it is possible to have a long range interaction. Furthermore, in more conventional scenarios for emergence of gravity-like theories, it is the nature of the emergent graviton as a Goldstone boson associated to the spontaneous breaking of Lorentz invariance to ensure the masslessness of the graviton.

Another interesting term is the vacuum term that we obtain in the ef\/fective Poisson equation~\eqref{eq:poisson}. It is generated by
purely quantum ef\/fects and it is related to the inequivalence between the Fock vacuum of the original bosons and the Fock vacuum of the quasiparticles, as its expression in terms of the Bogoliubov coef\/f\/icients makes manifest \cite{Finazzi, Girelli:2008gc}.
Interestingly, a~careful calculation shows that it is indeed a function of the so-called depletion factor, the ratio between the noncondensed phase and the condensed one. It is a measure of the quality of the condensate: the smaller it is, the fewer are the atoms that sit outside of the condensate.

The relevance of this term is not only its nature as a peculiar quantum vacuum ef\/fect, but also because it enters the ef\/fective
equations of motion \eqref{eq:poisson} exactly as the cosmological constant enters the Newtonian limit in general relativity.
Of course, this leads to the concrete possibility to discuss one of the most excruciating problems of modern physics, which is the explanation of the smallness of the observed cosmological constant.

Here, the smallness of the cosmological constant term would be explained by two basic observations \cite{Finazzi}: f\/irst of all, its origin is not that of a vacuum energy, as it is normally taken for granted in semiclassical gravity. Second, it is related to the depletion factor, which is in turn related to the very possibility of speaking about a condensate. This might be an alternative to the quantum f\/ield theory calculations of the vacuum energy, which are leading to the large discrepancy between the theoretical prediction and the observed value of the vacuum energy density \cite{CarrollCC,PaddyCC, WeinbergCC}.

Given that this model provides only an analogy, we cannot immediately export this lesson to the case of the real cosmological constant, but it certainly constitutes another evidence that the intuition of this term as a vacuum energy term might be fallacious, while more ref\/ined considerations, often related to thermodynamic or statistical behaviors \cite{GLCC2,GLCC1,VoloCC4,PaddyCC2,SorkinCC,VoloCC1,VoloCC3,VoloCC2}, might lead to an explanation of its origin and, ultimately, its magnitude.

\subsection{Multi-BEC, masslessness and the equivalence principle}
There are several points that make the case of a single BEC a rather poor analogy for the phenomena that we normally associate to gravitation. First of all, there is only one kind of particles, that make the model very poor on the matter side. Questioning the equivalence principle (in any form), or the emergence of Lorentz invariance for all the particle species is impossible in principle. Second, there is the important issue of the range of the ef\/fective Newtonian gravitational potential.

It is important, then, to extend the analysis as much as possible, as it has been done for kinematical analogues, by going from single component systems to many components.
This has been done in \cite{NBEC}, where the procedure that has been elucidated for the single BEC has been extended to the case of spinor BECs, where several components are present.

We are not going to review all the steps of the model, which can be found in the mentioned paper and do not add much to the discussion. Rather, we will focus on the outcome. The results are similar to the ones for the kinematical analogues. In absence of symmetries,
\begin{itemize}\itemsep=0pt
\item there is no Lorentz invariance at low energy, but rather multirefringence;
\item there is no obvious analogue gravitational potential that is singled out, but a rather complicated pattern of interactions with all the components of the condensate.
\end{itemize}

The ef\/fective model that emerges, in the most general case, does not represent a viable analogue for any sort of physically relevant regime of the gravitational interaction.
One might be pretty discouraged, at this point. However, interestingly enough, it is possible to cure all the diseases with one simple recipe: imposing a global symmetry on the system.

In a system with $N$ components, assuming an Hamiltonian which is invariant under permutations of the components among themselves, one can see that the following happens:
\begin{itemize}\itemsep=0pt
\item there is no Lorentz invariance, but f\/ields are organized in multiplets, within which a low energy (multiplet-dependent) Lorentz symmetry holds;
\item there is a single distinguished mode (associated to the simultaneous deformation of the condensate wavefunctions) which plays the role of the Newtonian gravitational f\/ield;
\item the analogue Poisson equation, arising from the system of Bogoliubov--De Gennes equations, does not contain a term leading to a short range potential: the analogue interaction is long range (or the graviton is massless);
\item the various kind of massive phonons are coupled in the same way to the analogue gravitational potential: the equivalence principle emerges.
\end{itemize}

Notice that, despite the very simple class of models that we are playing with, we are still able to
discuss plenty of interesting features of realistic gravitational dynamics, like the range of the interaction and the appearance of an equivalence principle.

This is teaching us an important lesson: in absence of strong symmetry arguments, the emergence of a theory as simple as Newtonian gravity can be hopeless\footnote{Obviously, one cannot hope to use the strategy of piling up all the matter f\/ields into a single (super-)multiplet of a ``grand unif\/ied" symmetry group. The general reason is that, while gauge bosons belong to the adjoint representation of the gauge group, the quarks and the leptons are in the fundamental one. This leads to severe constraints \cite{Distler} for the realization of a multiplet containing all the forms of matter f\/ields that we know, and hence for the opportunity to control Lorentz symmetry and the equivalence principle at once using this specif\/ic strategy.}. In order to cure all the possible diseases, symmetries have to be imposed on the microscopic model. This lesson is of course extendible from this simple class of toy models to more ref\/ined scenarios. Nonetheless, the BEC case can show basically all the potential dif\/f\/iculties that an emergent scenario has to address, besides getting the correct Einstein equations.

\section[Background Minkowski spacetime: the graviton as a Goldstone boson]{Background Minkowski spacetime:\\ the graviton as a Goldstone boson}\label{goldstoneboson}

The analysis of analogue models has highlighted a number of basic points that might lead to a failure, in the program of emergent gravity. The case of the BEC is instructive also at the dynamical level. However, it was essentially a nonrelativistic model. One might wonder what happens if we try to include special relativity in the game.

A simple point of view would be to consider gravitation as the interaction mediated by a helicity two massless particle moving over a background Minkowski spacetime. Therefore, before trying to emerge a form of background independent theory like general relativity, one might want to attempt to describe a model in which a helicity  two particle does propagate in a f\/lat background spacetime, allowing us to start to address the emergence of dif\/feo-invariance, albeit only in a perturbative setting. This is the subject of the present section.

\subsection{Comments on massless particles}

General arguments of quantum f\/ield theory already tell us that if we want such a particle to remain massless even in the presence of radiative corrections, it should be a Goldstone boson (if it is not a gauge boson). In the case of the multi-BEC the gravitational potential was a kind of Goldstone mode (even though it was associated to a discrete symmetry, not a continuous one) for an internal invariance group that is spontaneously broken by the appearance of the condensate. In absence of such symmetry breaking patterns, a custodial symmetry has to be introduced to protect the mass term of the f\/ield encoding the emergent graviton from large renormalization ef\/fects, allowing it to stay massless.

In addition to these caveats, we need to take care of some limitations that apply to Lorentz-invariant f\/ield theories containing massless particles. In particular, there is a well known theo\-rem\footnote{See also \cite{Case}.}, due to Weinberg and Witten \cite{WW}, which is often presented as a crucial (fatal, in fact~\mbox{\cite{Jenkins2, Jenkins1}}) obstruction for a successful emergent gravity program.
The theorem states precise limits for the existence of consistent theories with massless particles. It consists of two parts, and it states that, in a Lorentz invariant theory, conserved currents associated to global symmetries, as well as a covariantly conserved stress energy tensor, cannot contain, respectively, contributions from particles with spin larger than~$1/2$ and~1.
For a careful discussion of the proof of the theorem, and for further references, see~\cite{Loebbert}. For additional comments, see~\cite{Jenkins2, Jenkins1} and Section~7.6 of~\cite{AnalogueReview}.

The crucial ingredients for the proof of this theorem are $a)$ Lorentz invariance and $b)$ the nonvanishing charges obtained from Lorentz covariant vectors and tensors. Indeed, the proof of the theorem does not involve specif\/ic statements about the underlying Lagrangians, but only rather general symmetry considerations.

As any theorem, it hinges on specif\/ic assumptions, which are rather easily overcome. For instance, in case of the gauge bosons like the gluons and of the graviton the theory does not apply, since the current for the gluons is not Lorentz-covariant conserved, and the graviton does not possess a covariant stress-energy tensor (but rather a pseudo-tensor). In both cases, the conservation law involves the gauge covariant derivative.

This theorem, nonetheless, poses rather strong constraints on the possible theories that can be built in Minkowski spacetime. Of course one could say that gravity is not just the theory of a spin-2 particle in Minkowski spacetime (not all the manifolds coming from the solutions of general relativity are dif\/feomorphic to Minkowski spacetime). Nevertheless, it surely makes sense to consider the linearized theory in suf\/f\/iciently small neighborhoods. In this limit, then, the theorem does apply.

With this caveat in mind, we can say that in an emergent gravity program this theorem must be taken appropriately into account and appropriately evaded.
There are (at least) two ``obvious'' ways out:
\begin{itemize}\itemsep=0pt
\item in a setup where a nondynamical background is present, allow for Lorentz symmetry breaking, or
\item make the spacetime manifold and its geometry emerge as well.
\end{itemize}

The f\/irst option is rather straightforward, and it is essentially what could be pursued within scenarios like the one considered in analogue models, in which a preferred time function is specif\/ied. However, this is also a (conceptually high) price to pay: a step back from Minkowski spacetime to the notions of absolute space and time. Moreover, and most importantly, there is the issue of recovering a low energy approximate Lorentz invariance: as we have already seen, this is not at all an easy task without the requirement that the theory possesses some additional symmetries or requires f\/ine tunings which are able to realize this objective.

The second option is probably the most viable, conceptually appealing, but most demanding in terms of new concepts to be introduced. If no reference is made to a background Minkowski spacetime, but rather the graviton emerges in the same limit in which the manifold emerges, then there is no obvious conf\/lict with the WW theorem. Simply, what is called the gauge symmetry in terms of f\/ields living on the given spacetime is the manifestation of an underlying symmetry acting on the fundamental degrees of freedom, in the limit when they are reorganized in terms of a spacetime manifold and f\/ields (gauge f\/ields and gravitons in particular). We will be more specif\/ic on this special class of models in the second part of this paper.

The bottom line of this very concise overview of the WW theorem is clear: to obtain a~realistic model of emergent gravity one must ask for very special mechanisms to be at work in the model, {to bypass what would be rather compelling obstructions to the recovery of gravity even in the weak f\/ield limit}.

The WW theorem is part of a family of theorems that have been established on the limitations that
have to be respected to produce theories contained interacting f\/ields with high spins. However,
this topic is still under consideration: in the research area of higher spin f\/ield theories there is an ef\/fort to develop models of interacting f\/ield theories which are Lorentz invariant, gauge invariant, and containing spins higher than 2, thus avoiding the various no-go theorems on massless particles. For a pedagogical review, see \cite{Bekaert}.

\subsection{Gauge invariance from spontaneous Lorentz violation}
After this long preliminary discussion, we will brief\/ly describe the idea of evading the WW theorem by means of Lorentz violation. We will assume that there is a background Minkowski spacetime, and that there are some f\/ields def\/ined over it. As it has been discussed in the case
of Bjorken's model for emergent electrodynamics, it is possible that QED and the other gauge interactions can actually be simulated by non-gauge theories in which the mediators of the interactions are massless Nambu--Goldstone bosons. The idea is rather old \cite{Birula,Cho,Guralnik1,Guralnik2}, but in recent years the subject has been reconsidered \cite{Chkareuli:2001xe, Chkareuli:2001pd}.

In order for the Goldstone modes to have the correct coupling to conserved currents or to the stress energy tensor, they must have vectorial (or tensorial) nature. This implies that they should arise as the excitations around a ground state of a theory for a vector or for a tensor f\/ield possessing a potential term which allows for spontaneous symmetry breaking. The prototype of such a behavior is the bumblebee model \cite{Bluhm2, Bluhm1}, whose Lagrangian is
\begin{gather*}
 L = -\frac{1}{4} F_{\mu\nu}F^{\mu\nu} - \frac{\mu^2}{2} A_{\mu}A^{\mu} + \frac{\lambda}{4} (A_{\mu}A^{\mu})^4,
\end{gather*}
where $\mu$ has the dimensions of a mass and $\lambda$ is a dimensionless coupling. The classical
equations of motion read:
\begin{gather*}
 \partial_{\mu} F^{\mu\nu} +\mu^2 A^\nu + \lambda A_{\mu}A^{\mu} A^{\nu} = 0.
\end{gather*}
The ground state for such a theory has inf\/inite degeneracy, which can be labeled by constant vector f\/ields $\Aa^{\mu}$ satisfying
\begin{gather*}
 \Aa^{\mu}\Aa_{\mu} = -\frac{\mu^2}{\lambda}.
\end{gather*}
If we introduce a (timelike) constant vector f\/ield $n^{\mu}$ with unit norm and write $\Aa^{\mu} = M n^\mu$, we have that $M=\mu \lambda^{1/2}$.
These vacua break Lorentz symmetry, and hence we expect Nambu--Goldstone bosons to arise. Of course, the resulting theory seems to lead, almost unavoidably, to observable Lorentz violation.
In fact, there is more. In \cite{Chkareuli:2006yf} it has been shown that, if we tune the parameters of the model in such a way that, even though spontaneous symmetry breaking of Lorentz violation occurs, its ef\/fects (essentially in the scattering amplitudes predicted by the theory) are not visible, the model thus def\/ined displays also local gauge invariance as a~byproduct.

Concretely, this means that, when considering f\/luctuations around these backgrounds
\begin{gather*}
 A_{\mu} = M n^\mu + a_{\mu}
\end{gather*}
the polarizations of $a_{\mu}$ corresponding to the massless NG modes are coupled to the conserved current in such a way to mimick Lorentz invariant electrodynamics in a noncovariant gauge.

Technically, this can achieved whenever the $S$-matrix for the model has the properties of being transverse, i.e.\ for a process containing an external $a_\mu$ line with momentum $q^\mu$ and pola\-ri\-za\-tion~$\epsilon^{\mu}$,
\begin{gather*}
S(q, p) = \epsilon^{\mu}(q) {\mathcal M}_{\mu}(q,p),
\end{gather*}
where $p$ denotes collectively all the labels of the other states involved in the process, obeys
\begin{gather*}
{\mathcal M}_{\mu}(q,p) q^{\mu} = 0.
\end{gather*}
In this case, the longitudinal polarization of $a_{\mu}$, $\epsilon^{\mu}_{\rm longit} \propto q^{\mu}$,
ef\/fectively decouples from the scattering involving conserved currents. Of course, this implies that
a transformation
\begin{gather*}
\epsilon^{\mu}(q) \rightarrow \epsilon^{\mu}(q) + i\tilde{\alpha}(q) q^{\mu},
\end{gather*}
or, in real space
\begin{gather*}
a^{\mu}(x) \rightarrow a^{\mu} + \partial^{\mu} \alpha(x),
\end{gather*}
becomes a symmetry of the scattering amplitudes.

 Within this framework, spontaneously broken Lorentz symmetry, together with the \mbox{(un-)}ob\-ser\-vability criterion, leads to the emergence of local gauge invariance as an accidental symmetry (in the technical sense of the term, see \cite[Section~12.5]{WeinbergQFT1}).

This picture has been variously extended to the case of gravity \cite{Berezhiani,Bluhm2, Bluhm1,Nielsen,Jejelava,Potting,Kraus, Ohanian,Phillips}.
For this more exotic case involving tensor f\/ields, the background conf\/iguration is identif\/ied via conditions of the form
\begin{gather*}
{H_{\mu\nu}H^{\mu\nu}} = \pm M^{2},
\end{gather*}
where $M$ is again an unspecif\/ied model-dependent energy scale associated to the UV physics responsible for the ef\/fective model considered.
Clearly, the solutions to these equations do identify background conf\/igurations $A_{\mu}$ or $H_{\mu\nu}$ that, despite translational invariance, break a~number of generators of the Lorentz group. According to Goldstone's theorem, we should expect an appropriate number of
massless modes, associated to the number of broken generators.
Now, these Goldstone bosons can be coupled to the other f\/ields in such a way to mimic the pattern of gauge invariant f\/ield theories, in a way similar to the case discussed for the photon and the nonabelian gauge f\/ields.

Similarly to the case of the photon, in this case, to ensure the ef\/fective gauge invariance of the emergent graviton, i.e.\ the decoupling of the other polarizations, it is enough to ensure that the
scattering amplitudes involving external gravitons are transverse,
\begin{gather*}
S = \epsilon_{\mu\nu}(q) {\mathcal M}^{\mu\nu}(q,p), \qquad k_{\mu} {\mathcal M}^{\mu\nu} = 0.
\end{gather*}
Under this hypothesis, the scattering amplitudes of matter f\/ields (albeit not the Lagrangian) are invariant under the gauge transformations
\begin{gather*}
\epsilon_{\mu\nu}(q) \rightarrow \epsilon_{\mu\nu}(q) - i \big(k_{\mu} \tilde{\xi}_{\nu}+k_{\nu} \tilde{\xi}_{\mu}\big),
\end{gather*}
or, in real space,
\begin{gather*}
h_{\mu\nu}(x) \rightarrow h_{\mu\nu}(x)+ \left(
\partial_{\mu} \xi_{\nu}+\partial_{\nu} \xi_{\mu}
\right).
\end{gather*}
This ensures that the additional polarizations of the f\/ield $h_{\mu\nu}$ not associated to the helicity two graviton ef\/fectively decouple from the conserved stress energy tensor of the other particles, not participating to what we would call gravitational scattering. In this sense, then, one would have the emergence of dif\/feomorphism invariance at the linearized level.
For a discussion about the interplay of Lorentz invariance, gauge symmetry and scattering amplitudes, see the discussion in \cite{Weinberg:1964}.

\subsection{Limitations}

While the construction of these models show that it is certainly possible to generate at least at the classical level something that resembles gauge interactions (gravity included), there are a~number of questions that need to be addressed.

First of all, clearly, these results are perturbative in nature, being based on the expansion of the models around f\/ixed backgrounds, viewing the graviton as a linear perturbation around the given background
\begin{gather*}
 g_{\mu\nu} = \bar{g}_{\mu\nu} + h_{\mu\nu},
\end{gather*}
with gauge symmetry represented by the action of inf\/initesimal dif\/feomorphisms (i.e.\ involving the part of dif\/feomorphism group that is connected with the identity, and, in fact close to it in an appropriate sense),
\begin{gather}
 h_{\mu\nu} \rightarrow h_{\mu\nu} + \nabla_{\mu}\xi_{\nu}+\nabla_{\nu}\xi_{\mu} \label{eq:gaugegrav}
\end{gather}
with the only ef\/fect of Lorentz violation being equivalent a specif\/ication of a noncovariant gauge condition.

These model have to face the challenge of going beyond the linearized theory, including, for instance, the self-interactions of the gravitons with themselves in a way that is consistent, at low energies, with the expansion of the Einstein--Hilbert action (or more general actions).
This problem stands out even at the classical level.
At the quantum level, the challenge will be to reproduce the Ward identities, conservation of currents, and, in general, all the consequences of dif\/feomorphism invariance on the quantum theory. These will be the true test for these models.

A more concrete, worrisome problem is related to the percolation of Lorentz violating ef\/fects from high dimensional suppressed operators to low dimensional ones, a phenomenon that happens to be generic in quantum f\/ield theories with Lorentz violation. The results on the generation of gauge invariance from Lorentz breaking have to be confronted with the presence of radiative corrections to tree level results.
While it is certainly possible, by tunings, to resolve the issue, on general grounds, radiative corrections
will introduce visible ef\/fects even at low energies~\cite{Carroll, Kraus}. For instance, in the case of the graviton it can be argued that radiative corrections will make the dispersion relation depend on the background
\vev for the tensor f\/ield
\begin{gather*}
 k^{\mu}k_{\nu}\left( \eta_{\mu\nu} + a H_{\mu\nu}  \right)=0,
\end{gather*}
where $a$ is a computable coef\/f\/icient. The ef\/fect is visible at low energy, involving the part of the kinetic term with only two derivatives, and leads to direction dependent speed of propagation according to the symmetry properties of $H_{\mu\nu}$.
For extended discussion on the phenomenology of this class of models we refer to the literature \cite{Bluhm2,Carroll,Tasson,Kraus}.

However, the most serious drawback of these models is conceptual, and not practical: what is the origin and the dynamical principle giving origin to the background Minkowski spacetime? This latter appears to be a rigid structure which is not governed by equations of motion. Even though it is possible to have a special relativistic framework, clearly a general relativistic picture is incompatible with the presence of such a construction, which pinpoints preferred classes of reference frames/coordinate systems which are adapted to the presence of the background.

\section{Towards pregeometry}\label{diffeo}

The models presented so far, while highlighting a number of interesting possibilities, are quite unsatisfactory when confronted with the pillars of general relativity (or its extensions).
One of the most distinguished features of the latter is the invariance of the physical content of the theory under the action of dif\/feomorphisms. There are two senses in which this is relevant. A more conceptual one, which is related to the fact that a theory should not make reference to rigid background scaf\/foldings, when dealing with its predictions, whose nature should be essentially relational.
The second reason is more practical, as it strongly inf\/luences also the dynamics of the theory itself. Indeed, loosely speaking, one can view general relativity as a sort of gauge theory for the dif\/feomorphism group, and the invariance under this group dictates the way in which the various matter f\/ields should be coupled to geometry and among themselves.

In fact, dif\/feomorphism invariance can be shown to lead also to a specif\/ic restriction of the form of the dynamical equations for the gravitational f\/ield, which has to be governed by constraints obeying essentially a Dirac algebra \cite{Kuchar}.

In more mundane terms, this means that the gravitational f\/ield, when expanded around f\/lat spacetime, can be described in terms of a Pauli--Fierz Lagrangian for a helicity two particle, coupled to the SET of the other f\/ields. Furthermore, dif\/feomorphism invariance selects also the possible continuations to the nonlinear regime (given that the action should be specif\/ied in a~dif\/feomorphism invariant way).

Interestingly enough, dif\/feomorphism invariance is also another way to evade the WW theo\-rem, given that, for the graviton, it is not possible to def\/ine a (Lorentz invariant) covariantly conserved stress energy tensor, the energy of the gravitational f\/ield being contained in a pseudo tensor.

\subsection{Spin systems and quantum graphity}

It is worth to mention some approaches where the graviton emerges as a collective mode exploi\-ting certain forms of phase transitions which do not make use of order parameters, thus promising to give rise to dif\/feomorphism invariance not through conventional symmetry breaking.

As we have said earlier, one could interpret the emergence of dif\/feomorphism invariance as the approximate validity of the Hamiltonian constraints. In this sense one might be tempted to construct an emergent model for quantum gravity for which the
constraint surface of the classical phase space is just the location of the minima of some very steep potential.
If this happens, there is room in the model to obtain a helicity two mode that can be formally described by
a rank two symmetric tensor f\/ield with gauge symmetry given by \eqref{eq:gaugegrav}.
We have already reviewed some of the approaches supporting this idea. Here we want to brief\/ly mention some recent attempts to achieve this by means of suitably designed lattice/spin systems.

The models of emergent gauge symmetries that we were discussing previously are all based on the exploitation of the f\/ield theoretical properties of phase transitions characterizable in terms of Landau--Ginzburg theories for certain tensorial order parameters. However, these do not exhaust the list of possible phase transitions. Recently, it has been realized that there are interesting (quantum) phase transitions that cannot be characterized by order parameters associated to local f\/ield theories. Rather, they are characterized by topological properties of the ground state of the model. They are part of the subject known as topological order \cite{Wenbook}.

Within this context, a prominent role is played by local bosonic models: bosons sitting on the nodes of lattices, whose dynamics is described by certain properly designed Hamiltonians. It can be shown that, under certain conditions, the models undergo specif\/ic forms of phase transitions, called string-net condensation, whose results are rather striking. Indeed, the analysis of the excitations above the ground states shows that they are indeed described in terms of fermions interacting via an emergent $U(1)$ gauge interaction \cite{levinwenrmp2005,wen2003b, levinwenprb2006, wen2003}.

This is certainly an important result that is undergoing rather intense investigations. Interestingly enough, concepts of topological order, as well as of string net condensations have been shown to bear strong resemblance with spinfoam models for quantum gravity \cite{bianca, spinfoams}.
Besides this, local bosonic models have been considered to understand whether it is possible to emerge the graviton, within this class of models, as a collective excitation around a string-net condensed state.
To this purpose, it has been shown that it might be possible to design Hamiltonians leading to the desired result
\cite{GuWen1,GuWen2}. We refer to these papers for the technical details of the construction of the models and the methods to extract the physical content.

Despite the obvious interest of this outcome, there are some basic dif\/f\/iculties that, to date, have not been overcome.
First of all, these models are based on f\/ixed nondynamical lattices, thus leading to worrisome features that require some additional care.
The f\/irst practical inconvenience is that a f\/ixed nondynamical lattice for sure breaks Lorentz symmetry. Therefore, as we have already discussed, the recovery of a low energy notion of Lorentz invariance will require the imposition of a custodial symmetry, or, in the worst case, a f\/ine tuning in the Hamiltonian. Similarly, the presence of a Hamiltonian might suggest that there is also a preferred notion of time, i.e.\ the lapse function has a f\/ixed value and does not represent a Lagrange multiplier associated to the Hamiltonian constraint of reparametrization invariant theories like general relativity.
Finally, the background nondynamical lattice constitutes a violation of dif\/feomorphism invariance/background independence in the hardest possible way. Therefore, extra polarizations of the graviton have to be expected and have
to be made massive by hand, with a mass scale whose origin is not clear.

A way to attack these problems is suggested by known examples of models of quantum gravity with fundamental discreteness, like (causal) dynamical triangulations \cite{CDT2,CDT1,CDT4,CDT3, CDT0}, causal sets \cite{causets,henson}, and considerations within loop quantum gravity \cite{RovelliBook}, spinfoam models \cite{spinfoams} and group f\/ield theories \cite{GFT1,GFT3,GFT2,GFT4}. All these approaches indicate that an obvious way out should be a dynamical random lattice.
While the ``dynamical'' part is still essentially work in progress, on which we will comment later, the
random nature of the lattice might be able to save Lorentz (or Poincar\'e) invariance, at least in a statistical sense, without leading to preferred frame ef\/fects or modif\/ied dispersion relations \cite{Sorkin}.

An attempt in this direction which is treatable with available techniques is represented by the so-called quantum graphity models \cite{qgraph3,qgraph2, qgraph1} (see also \cite{onion,caravelli,Conrady, Hamma} for related work). The idea is very simple. Instead of starting with a given f\/inite lattice with N nodes, which consists of sites and links connecting the sites, one can imagine to start with its progenitor, the complete graph with~$N$ nodes. The lattice structure emerges dynamically when an appropriate Hilbert space is associated to this graph, in which each state corresponds to a particular connectivity conf\/igurations in which links are turned on or of\/f. Some of these conf\/igurations might be interpreted in terms of regular lattices in certain dimensions (which is variable, in these models), while others may not have this interpretation.

The Hamiltonian, and the partition function describing the statistical mechanics of the system, are constructed in such a way to obtain a macroscopic notion of space (and of its geometry) only in a statistical sense.
In particular, the investigation of the statistical mechanical properties of the models tends to show the existence of at least two phases: a high temperature phase in which the state of the system is dominated by graph conf\/igurations which are highly connected and do not have an interpretation of a nice smooth geometry, and a low temperature phase in which it is possible to recognize the pattern of smooth regular geometry. The phase transition separating the two phases has been called geometrogenesys.

While numerical methods are the main road to the content of these models, a mean f\/ield approximation technique has been developed \cite{caravelli}, which makes use of a mapping of the models to particular spin systems similar to the Ising model, of\/fering the possibility of an analytic study, within the given approximation scheme.

Of course, establishing the existence of a phase in which a macroscopic notion of geometry can be given is only part of the problem. The emergence of the right form of gravitational interaction is still work in progress. However,
interestingly enough, it is possible to see that, in certain cases \cite{onion}, these models do possess states associated to
peculiar graphs in which the excitations do propagate as they would do in a continuum spacetime geometry with a trapping region (ideally, a black hole).

These kind of models are very intriguing, but they are not free of ambiguities. First of all, the construction of the Hilbert space f\/irst, and then of the Hamiltonian suf\/fers from ambiguities. Lacking a notion of geometry with which operators can be classif\/ied into UV or IR, a systematic construction of the Hamiltonian following the rules of ef\/fective f\/ield theories appears to be impossible. Given the fact that a complete graph is used at the foundation implies that the emerging geometry could possess any dimension (actually, any number from~0 to~$N$): this is certainly an ambiguity that requires some care. Finally, given that the model is presented with a proper Hamiltonian, the status of Lorentz and time reparametrization is unclear: Lorentz invariance violation might be present, the long range continuum limit being similar to Lifshitz-type theories~\cite{Horava}, or just be a spurious ef\/fect due to the fact that the def\/inition of the model corresponds to the choice of a not covariant gauge.
Apart from the dimensionality problem, many of these issues are af\/fecting the class of random models that we presently know: for instance, the imposition of causality onto random matrix models \cite{dariohenson} seems
to point in the direction of having preferred foliation ef\/fects, or, at least, nondynamical structures.

\subsection{Emerging dif\/feomorphisms}

At this point it is clear that the next kind of questions that we need to address, in our path for the construction of a theory of emergent gravity, is whether dif\/feomorphism invariance can be made to emerge as well. There are two scenarios in which this question can be addressed.

The f\/irst one is obvious, and will be the subject of the next sections: there is no spacetime manifold, or, at most, a nondynamical topological manifold over which no metric or connection structures are specif\/ied. In this case dif\/feomorphism invariance should be guaranteed by the absence of background absolute structures \cite{Giulini}, which forces the observables and the dynamical structures to be purely relational. We will discuss some of these models in the next section.

The second scenario is the one in which there is a background spacetime with a given geometrical structure, nondynamical, providing a scaf\/folding, and dif\/feomorphism invariance is just an emergent concept, valid only in a certain corner of the phase space of the theory at hand.

Essentially, the key idea that is behind emergent dif\/feomorphism invariance has already become apparent in the case of the graviton as a Goldstone boson. Dif\/feomorphism invariance can be considered to have emerged whenever the form of the theory at hand can be made to coincide with a gauge f\/ixed version of dif\/feomorphism invariant theories in the same regime.

Concretely, one can picture the emergence of gauge invariance in a specif\/ic corner of the phase space of a theory as the fact that, in correspondence of that specif\/ic regime, the dynamical equations of motion contain a potential term which becomes so steep to be able to approximate, for all the practical purpose, a Dirac delta enforcing the constraints characterizing the canonical description of a true gauge theory.

Notice that there is a dif\/ference with the St\"uckelberg method of introducing auxiliary f\/ields to make a theory look gauge invariant. Here we are making a statement about the physical content of the theory, having in mind the direct comparison of the physical quantities as computed in the two frameworks. Emergence of a dif\/feomorphism invariant theory would be then qualif\/ied as the property of all the computable physical quantities to be in agreement (within approximations) with those computed, in the appropriate regime corresponding to the specif\/ied background, in a dif\/feomorphism invariant theory.

Concretely, if all the processes of a given model, like the scatterings mediated by what would be identif\/ied as the mediator of the gravitational interaction, do reproduce (exactly or approximately) the same results obtained within dif\/feomorphism invariant models, expanded around the same background def\/ining the regime of the theory that has to be compared to, the empirical content of the two theories can and has to be considered equivalent (see, on this, \cite{BarceloJannes} and references therein for the discussion of the case of Lorentz invariance).

Apart from the case of the graviton as a Goldstone boson scenarios, discussed in the previous section, there is only the case, to the best of our knowledge, studied in \cite{Nordstrom}, of the emergence of dif\/feo-invariant Nordstr\"om theory for conformally f\/lat gravity (and of Lorentzian signature) out of a theory def\/ined in a f\/lat Euclidean background.
Obviously, such studies do make sense only in a perturbative, semiclassical regime, for which a notion of spacetime manifold, with its geometry, can be specif\/ied. A comparison with the full quantum gravity regime of background independent theories would not be appropriate.

Of course, the emergence of dif\/feomorphism invariance, in these regimes, would suf\/fer of the same problems of the emergent low energy Lorentz invariance: being an approximate symmetry, it will be possible to discover specif\/ic signatures of the violation of the conservation equation associated to it.

There are at least two kind of ef\/fects that one has to expect. First of all, given that dif\/feomorphism invariance implies the covariant conservation of the SET, $\nabla_{\mu}T^{\mu\nu}=0$, the most immediate place to look for would be the total matter energy balance in gravitational processes, looking for a nonconservation of the SET
\begin{gather*}
\nabla_{\mu}T^{\mu\nu} \neq 0,
\end{gather*}
albeit this line of thought might be leading towards a confusion with the problem of dark matter (or more general dark components of the matter content).

Another obvious direction to be pursued would be the detection of extra modes for the gravitational f\/ield. In linearized general relativity, local Lorentz invariance and dif\/feomorphism invariance reduce the 10 components of the metric tensor to two propagating degrees of freedom, describing gravitons. The other components are essentially unphysical gauge modes. However, a breaking of either Lorentz invariance (which accompanies these models as well, generally) or dif\/feomorphism invariance would promote some of the would-be gauge modes to the status of physical modes.
These modes would be extremely dangerous for the compatibility of the theory with the observations, given that, on general grounds, their coupling to the matter SET would be as strong as the one of the emergent gravitons. However, the steepness of the potential might be of help in making these extra polarizations extremely massive, so that it will be extremely dif\/f\/icult to excite them.
While this is certainly a solution, it opens a naturalness problem related to the magnitude of the energy scale $M$ controlling the
extra polarizations to the other scales of gravity and particle physics.

Of course, quantum mechanics has to be appropriately taken into account. In addition to the previously mentioned problems, radiative corrections might lead to dangerous corrections which would make the models not viable. For instance, the mass term for the would be graviton could be subject to large renormalization ef\/fects, given that the symmetry that would able to protect it, dif\/feomorphism invariance, is not present. More work is needed, in this direction. A~f\/irst attempt to def\/ine an ef\/fective f\/ield theory which would be able to systematically address the phenomenology of the breaking of dif\/feomorphism invariance is presented in
\cite{Donoghue}, to which we refer for further readings.

For all these reasons, models of emergent gravity with a background metric structure have to be considered with particular care, and might not be considered as completely satisfactory and physically viable models of emergent gravity.

\section{Background independent pregeometrical models}\label{pregeometry}

The discussion has brought us to the last step: emerging geometry from pregeometrical degrees of freedom. It is often stated that, on general grounds given by considerations of classical and quantum f\/ield theories def\/ined over Minkowski spacetime (e.g.\ the Weinberg--Witten theorem), that a model of emergent general relativity will generically bear signs of Lorentz symmetry violation and dif\/feomorphism symmetry breaking.

Of course, this is a simplistic point of view due to the particular starting point (that might be circumvented, as we have discussed), that is f\/ield theories on some form of metric background.
There are very general dif\/f\/iculties in def\/ining working models for emergent gravity which do not make reference to such background structures, but this does not means that there are no models at all.

To def\/ine dif\/feomorphism invariant quantum f\/ield theories it is necessary to def\/ine appropriately a volume element with which to def\/ine, in an invariant way, an integration over the (background) manifold\footnote{Here the manifold might be considered as given or not. The important point is that it is just a topological manifold, without any preassigned metric structure.}. The role of the volume element, in the case of general relativity, is played by the square root of the modulus of determinant of the metric (or by the determinant of the vielbein in a f\/irst order treatment).
In absence of these structures, something else has to be proposed\footnote{Interestingly, in Schr\"odinger book \cite{Schroedinger} on general relativity it is mentioned the possibility of a purely f\/irst order action for pure gravity that does not make reference to a metric, but only to a connection. The action reads
\begin{gather*}
 \int d^4x\left(\det(R_{\mu\nu}(\Gamma))\right)^{1/2}.
\end{gather*}
Of course, the role of the metric is not just to provide us a notion of connection and curvature, but also to def\/ine the kinetic terms for the matter f\/ields. It is not clear how to couple matter to this kind of models.}.

The models we are going to describe (which are related one another) are formulated in such a~way to provide, f\/irst of all, a proper def\/inition of volume element. Then, the metric is obtained as a collective f\/ield, with its dynamics induced along the lines proposed by Sakharov. However, they are not merely a form of induced gravity, where the metric tensor only acquires a dynamical term. Rather, they promise to deliver the metric tensor itself.

The class of models we are going to discuss have been f\/irst formulated in a series of papers \cite{Akama2,Akama4,Akama1,Akama0,Akama3}, in which Sakharov's idea is implemented in a model where no explicit notion of geometry is introduced at the beginning. Rather, the metric tensor (or the vielbein/vierbein in a D/4 dimensional f\/irst order formalism) is interpreted as arising from a nonvanishing vacuum expectation value of certain composite operators
def\/ined in dif\/feomorphism invariant quantum f\/ield theories. In the following we are going to discuss the models at length, given that they represent interesting and signif\/icant alternatives to the scenarios that we have described so far, being able to be pregeometric without losing contact with the language and the concepts of quantum f\/ield theory.

Consider a multiplet of scalar f\/ields $\phi^i$, $i=1,\dots,N$, whose classical action is given as
\begin{gather}
 S[\phi] = \int d^4 x \left(-\det\left(  \partial_{\mu}\phi^i\partial_{\nu}\phi^j\delta_{ij} \right)\right)^{1/2} F(\phi), \label{eq:NG}
\end{gather}
where $F$ is an arbitrary potential.
Despite this action might appear to be rather exotic, in two spacetime dimensions is just the Nambu--Goto action for the string embedded into an $N$-dimensional spacetime, with embedding functions given by the f\/ields~$\phi^i$. In higher dimensions, it represents the action of a membrane embedded in a pseudo-Riemannian space.

A partition function for this kind of action is easily written
\begin{gather*}
 Z = \int \DF [\phi] \exp(-iS[\phi]).
\end{gather*}

Clearly, such a partition function is dif\/f\/icult to be def\/ined rigorously, at least in the bosonic case. The general idea to get around this dif\/f\/iculty is to use a semiclassical treatment. Indeed, at the classical level, the Nambu--Goto-like action \eqref{eq:NG} can be translated into a Polyakov-like action with an auxiliary metric
\begin{gather*}
S'[\phi] = \int d^4 x \left(-\det{g^{\mu\nu}} \right)^{-1/2} \left( -\frac{1}{2} g^{\mu\nu}
\partial_{\mu}\phi^i\partial_{\nu}\phi^j\delta_{ij} - F^{-1}(\phi) \right),
\end{gather*}
where we are using the auxiliary f\/ield $g^{\mu\nu}$.

Notice that this auxiliary f\/ield appears algebraically, and hence plays the role of a Lagrange multiplier. The equations of motion derived from the variation of this classical action lead to
\begin{gather*}
 g_{\mu\nu} = \partial_{\mu}\phi^i\partial_{\nu}\phi^j\delta_{ij} F(\phi).
\end{gather*}
Using this equation to eliminate $g^{\mu\nu}$ from $S'$ leads to the action $S$.

Clearly, this mapping holds at the classical level. A full quantum theory will presumably forbid this nice reparametrization of the model. The general idea is that, in the semiclassical limit of the model, the metric tensor arises as the \vev of a composite f\/ield
\begin{gather*}
 g_{\mu\nu} = \langle \partial_{\mu}\phi^i\partial_{\nu}\phi^j\delta_{ij} F(\phi)\rangle.
\end{gather*}

A semiclassical analysis, following Sakharov's idea, leads to the conclusion that the dynamic of this f\/ield will be
generated through radiative corrections. In particular, the divergent part of the one loop ef\/fective action contains the Lagrangian density
\begin{gather*}
 L_{div} = \frac{N}{(4\pi)^2} \sqrt{-g} \left( \frac{\Lambda^4}{8} + \frac{\Lambda^2}{24} R(g)\right),
\end{gather*}
where $N$ is the number of scalars, $R(g)$ is the Ricci scalar derived from the metric $g$ and $\Lambda$ is an energy scale associated to a cutof\/f to be introduced to regulate the theory.

The scalar model presents a number of issues, in its foundations, that suggest that it should be taken as a provisional model. The very def\/inition of a path integral for such a theory, notion of stability, the nondegeneracy of the metric etc are not addressed at all.

Part of these concerns can be mitigated if fermions are used instead of bosons. If fermions are included, the starting point is the action
\begin{gather}\label{eq:NGfermion}
 S[\psi,\bar{\psi}] = \int d^4x \left( \frac{i}{2} \bar{\psi}^i \gamma^{a} \partial_{\mu} \psi^j \delta_{ij} \right)F(\psi,\bar{\psi}),
\end{gather}
where we are introducing Dirac's gamma matrices $\gamma^a$. Notice that these last indices are Lo\-rentz/in\-ter\-nal indices, and not vector ones, and refer to the fact that the fermions transform nontrivially under a global Lorentz transformation acting on these indices. Notice also that there are additional internal indices $i,j=1,\dots,N$ which manifests the fact that, in order to have a nonzero result for a partition function, several fermionic f\/ields are required.

In such a model, besides the manifold, one has to postulate the existence of a spin structure. This is not yet the specif\/ication of a metric, given that the def\/inition of a metric tensor still requires the vielbein.
In other terms, in a model with fermions one is forced to introduce a~notion of internal metric, acting on internal/Lorentz indices. A deeper analysis of the situation can be found in \cite{Percacci}.

Given the Grassmannian nature of the fermionic f\/ields, the def\/inition of the quantum partition function is less problematic than the corresponding bosonic case.
To extract the content of the model it is possible to perform a semiclassical analysis. At a semiclassical level,
the action~\eqref{eq:NGfermion} is equivalent to the following action
\begin{gather*}
 S'[\psi,\bar{\psi}] = \det(e^a_{\mu})^{-1} \big( e^{\mu}_{a} \bar{\psi}^i \gamma^{a} \partial_{\mu} \psi^j -3 F^{-1/2} \big).
\end{gather*}
In analogy with the scalar case, the vielbein arises as the expectation value of a composite f\/ield, a fermion bilinear
\begin{gather*}
 e^{a}_{\mu} = \frac{i}{2} \langle \bar{\psi}^i \gamma^{a} \partial_{\mu} \psi^j  F^{1/3} \rangle.
\end{gather*}
Interestingly enough, the action~\eqref{eq:NGfermion} is not invariant under local Lorentz transformations, since, to do so, one should introduce a spin connection, which is of course forbidden by the rules of emergent gravity.
The way in which spin connections, and Yang--Mills gauge f\/ields are introduced in this model is through the potential term~$F$. Shaping it in such a way to mimick the four Fermion interaction terms used in models {\'a la} Bjorken, we can generate spin connection and Yang--Mills gauge f\/ields as composite f\/ields.

Indeed, def\/ining
\begin{gather*}
 -3 F^{-1/3} = \alpha \big(\bar{\psi}^i \big\{\gamma^a,[\gamma^b,\gamma^c]\big\}\psi^i\big)^2,
\end{gather*}
then the model, at the semiclassical level, becomes equivalent to a local Lorentz invariant model where the
spin connection is determined in terms of the composite f\/ield
\begin{gather*}
 A^{abc} = 16 \alpha \big{\langle} \bar{\psi}^i \big\{\gamma^a,[\gamma^b,\gamma^c]\big\}\psi^i \big{\rangle}.
\end{gather*}

The application of heat kernel techniques to the computation of the one loop ef\/fective action has to take into account the fact that the background is a metric af\/f\/ine one. This generates radiative corrections that contain, besides the Einstein--Hilbert and the cosmological constant term, a potential term for the torsion. However, the torsion tensor appears without a kinetic term, and hence it is not propagating. For further discussions, see~\cite{TerazawaActions}.
Similar ideas can be used to def\/ine Yang--Mills f\/ields and their dynamics.

Summarizing, these are emergent gravity models, which are locally Lorentz, gauge and dif\/feomorphism invariant. This is achieved by the combination of ideas of composite gauge bosons, semiclassical analysis and quantum f\/ield theory that we started from.

These ideas have been developed in various directions \cite{Amati:1981tu,Amati:1981rf,Denardo1,Denardo4,Denardo2,Denardo3,Denardo0, Yoshimoto}. In \cite{Amati:1981tu,Amati:1981rf} it has been presented a model that addresses the spontaneous generation of a mass scale (crucial to separate UV and IR degrees of freedom) and, simultaneously, the possibility of unif\/ication of Yang--Mills and gravity at high energy. More recently, Wetterich \cite{Hebecker,Wetterich1,Wetterich2,Wetterichtime} has proposed to use fermionic models to generate gravity along the same lines.

These models are very promising. However, they have to face serious dif\/f\/iculties. Let us brief\/ly mention them.
First of all, one might be worried that a gauge invariant theory might lead to nonvanishing vacuum expectation values of gauge variant operators, as dictated by Elitzur's theorem \cite{Elitzur}.
However, this dif\/f\/iculty is only in the language, and not in the substance. Indeed, the presence of a nontrivial \vev of a metric tensor or of the vielbein might be expressed in gauge invariant terms by the equivalent statement that gauge invariant geometrical quantities (areas, length, angles, volumes) do acquire nonvanishing \vev s. The logic is the same behind the gauge invariant formulation of spontaneous symmetry breaking of gauge f\/ield theories with a Higgs mechanism \cite{oraifeartaigh,struyve}.

A more urgent problem is the very def\/inition of the path integrals. However, while for bosons the model might seem to
be too contrived, for fermions the situation improves considerably and one might imagine that an appropriate def\/inition might be provided, once the divergences are appropriately tamed.

A serious problem concerns the conditions under which nonvanishing \vev s for geometrical operators can be generated.
In QFT in ordinary Minkowski spacetime, this is controlled by the specif\/ication of the potential of the theory, that in turns determines the properties of the ground state. The ground state is def\/ined as the state that minimizes the energy, which, in turn, is def\/ined to be the charge conjugated to a globally def\/ined timelike Killing vector f\/ield.
In pregeometry there is no such a thing as a metric: how is the ground state def\/ined, then? How can we make statements like
``at low energy the ef\/fective f\/ield theory is~$X$'', if we do not have a notion of energy? In this sense, the emergence of scales has to be addressed, and, after this issue has been settled, one can imagine to study the RG f\/low of the theory with the methods that are being developed in these years, especially in the context of asymptotic safety (see the contribution in this special issue and also~\cite{percacciemergent} for a discussion of the relevance for the emergent gravity approach).

However, the most serious problem is the fact that, for these semiclassical arguments to work, the composite metric or the composite vielbein is {\it assumed} to be invertible. This is dif\/f\/icult to explain from a dynamical perspective (see~\cite{Percacci} for a discussion of this point). It might be that the ultimate reason is some form of Coleman--Weinberg ef\/fect in quantum gravity~\cite{PercacciCW}, in which the relevant scales and a potential leading to  a nondegenerate metric as a ground state conf\/iguration are generated dynamically by radiative corrections. In absence of such a~dynamical explanation, the only option is to continue to work in a kind of self-consistent mean f\/ield approximation \cite{PercacciFloreaniniMFA, Percacci}.

It seems that the subject is not suf\/f\/iciently explored. There is a lot of potential for phenomenology given the close contact with the language and the tools of quantum f\/ield theories, for which a vast number of techniques are available. In addition, there is an obvious relevance for fundamental physics since this class of models can be linked very easily to the issues of unif\/ication of all the interactions. Finally, it seems natural that, in such a scenario, the cosmological singularities of general relativity can be replaced by geometrogenesis phase transitions, as it is been conjectured on the ground of the experience with quantum graphity models.

\section{Lorentz invariance reloaded: emergent signature}\label{signature}

Lorentz symmetry plays an important role, in gravitational theory, both at the kinematical and at the dynamical level. However, in an emergent scenario, it is not obvious that Lorentz symmetry will be implemented at the fundamental level, as we have seen.

The situation might be that Lorentz invariance is an infrared symmetry, with the ultraviolet theory being essentially dif\/ferent (see, for instance, anisotropic scaling scenarios \cite{Horava}). The situation might be in fact much more drastic. Indeed, the very notion of signature might be only an emergent one, with the background being more similar to a timeless Euclidean theory rather than Lorentzian.

Scenarios might be conceived in which the internal Lorentzian metric $\eta_{ab}$, in a f\/irst order formalism in which the metric tensor is derived from tetrads $g_{\mu\nu} = e^{a}_{\mu}e^{b}_{\nu} \eta_{ab}$, is just a specif\/ic \vev of another f\/ield (a transforming as a tensor in the internal indices). This is the logic behind the models in which gravity is formulated in terms of $GL(4)$ gauge theories \cite{Percacci}.
In this picture, the selection of Lorentzian rather than Euclidean signatures is a dynamical problem to be addressed in terms of spontaneous symmetry breaking.

Notice that this is conceptually very dif\/ferent from Wick rotation. A Wick rotation is a procedure of analytic continuation of a theory from Lorentzian to Euclidean signature by analytically continue the time coordinate to the complex plane, and hence restricting it to the imaginary axis. The case for dynamical signature is completely dif\/ferent: no analytic continuation is performed. Rather, the theory allows for Lorentzian, Euclidean or more general signatures, at the level of the equations of motion. The signature is then selected dynamically.

In pregeometrical models like those described in the previous section the situation is simi\-lar~\cite{Wetterichtime}. The theory is typically formulated in terms of f\/ield theories with a large global  symmetry group, and then the signature is selected by the particular symmetry breaking pattern that accompanies the formation of metrics or tetrads as composite f\/ields.

This is not the only point of view. In \cite{greensite0} it has been proposed to parametrize the internal metric as
\begin{gather*}
 \eta_{ab} = \mathrm{diag}(e^{i\theta},+1,+1,\dots,+1),
\end{gather*}
which then smoothly interpolates between the Riemannian and the Lorentzian signatures, keeping the metric nondegenerate (but complex, in general) for any value of $\theta$. This parameter can inherit its own dynamics from one loop corrections generated by matter f\/ields. It is possible to compute a form of (complex) ef\/fective potential
which of\/fers the possibility of computing the expectation value for the f\/ields itself.

By analyzing the contribution of fermionic and bosonic f\/ields, it is possible to show that the model can be made consistent only in four spacetime dimensions, by adjusting the matter content. In turns, the analysis of the potential also predicts Lorentzian signature. Similar ideas concerning the dynamical generation of signature can be found in \cite{greensite2,greensite1,elizalde1994,hayward1995}.

There is a third way to start from a theory which is fundamentally Euclidean but that manifests Lorentzian ef\/fective metrics (and hence propagation of signals). The idea was put forward in \cite{Nordstrom} and it is based on the observation that, in a second order PDE, the metric tensor is identif\/ied with (the inverse of) the matrix which is multiplying the second order derivatives:
\begin{gather*}
 h^{\mu\nu} \partial_{\mu}\partial_{\nu} \phi + a^{\nu} \partial_{\nu} \phi + f(\phi) = 0.
\end{gather*}
The signature of the tensor $h^{\mu\nu}$ determines whether the PDE is hyperbolic, parabolic or elliptic. In wave equations encountered normally in physics, the tensor $h^{\mu\nu}$ is an externally determined tensor, independent from the f\/ield that is propagating over it. However, when considering (scalar) f\/ield theories which are characterized by (f\/irst order!) nonquadratic kinetic terms, the situation is more complicated. In particular, in a system with Lagrangian
\begin{gather*}
L = F(X),\qquad X= \gamma^{\mu\nu} \partial_{\mu}\phi\partial_{\nu}\phi,
\end{gather*}
with $\gamma_{\mu\nu}$ a nondegenerate background metric, the principal symbol of the equation of motion is given by
\begin{gather*}
h^{\mu\nu} = \frac{\partial F}{\partial X} \gamma^{\mu\nu} +2\frac{\partial^2 F}{\partial X^2}\gamma^{\mu \rho}\partial_{\rho}\phi\gamma^{\nu \sigma}\partial_{\sigma}\phi.
\end{gather*}
This tensor shows an explicit dependence on the derivatives of $\phi$. In turn, this can lead to the presence of solutions for which the tensor $h^{\mu\nu}$ is Lorentzian even when $\gamma_{\mu\nu}$ is Euclidean. We refer to~\cite{Nordstrom} for the details of how this can be used to generate a dif\/feo-invariant emergent Nordstr\"om theory of gravity in a Euclidean f\/lat background. See also~\cite{tesi} for a careful discussion of the role of the various symmetries that are needed to achieve the result.

Touching a foundational notion of physics as time does not come without a huge price to be payed. First of all,
in models in which Lorentz invariance is emerging from a Euclidean theory as the last models described, Lorentz symmetry breaking is unavoidable. In models in which the fundamental symmetry group is larger, like $GL(4)$, or in pregeometrical models described in \cite{Wetterichtime}, the selection of a specif\/ic signature via spontaneous symmetry breaking leads to the appearance of a Nambu--Goldstone mode for each of the generators associated to the broken symmetries. Approaches as the ones proposed by Greensite are associated to the appearance of complex actions: there the status of unitarity has to be kept under control.

These problems are just the tip of the iceberg. Indeed, without a def\/inition of Lorentzian metric, we cannot give a def\/inition of energy and hence of ground state. The classif\/ication of states in terms of energy loses its meaning, and, as a consequence, the selection of states in terms of energetic considerations (stability, essentially) is not well def\/ined.
Furthermore, while any locally paracompact dif\/ferentiable manifold can be endowed with a Riemannian structure, it is not true that we can put a Lorentzian structure over it. For this to be the case, the manifold should admit the existence of a globally def\/ined, nowhere vanishing smooth vector f\/ield. This is a rather strong topological restriction, which refers to the pregeometric structure.

The emergence of time seems to be a rather exotic curiosity, leading to all sorts of paradoxes. However, it is a possibility that has to be considered very carefully, since the simplest low energy theory at our disposal to describe gravity, general relativity, does possess a dynamical content that seems to be intimately related to the presence of light cones, i.e.\ to the Lorentzian signature of the metric tensor. This is clearly purported by the derivations of Einstein equations by means of thermodynamical considerations on { local horizons}.

\section{Challenges and future directions}\label{challenges}
Hopefully this overview has given at least the f\/lavor of various attempts to describe gravity as an emergent {dynamical} phenomenon, their merits as well as the problems that they raise. The proposals cover a rather wide range of dif\/ferent points of view and meet dif\/ferent amount of success and dif\/ferent kinds of shortcomings.
We have not mentioned other possibilities, most notably a specif\/ic kind of matrix models for higher dimensional gravity (see~\cite{Steinacker} and references therein), matrix models for 2D quantum gravity~\cite{MM2, MM1} and their extensions in term of tensor models~\cite{Razvan} and group f\/ield theories \cite{GFT1,GFT3,GFT2,GFT4}. These models are extremely interesting given that their nature is essentially algebraic or combinatoric, and the notions of continuous dif\/ferential geometries will emerge in certain specif\/ic regimes.

In a certain sense, also the AdS/CFT correspondence is an example of emergent gravity if seen from the CFT side.
However, while on these topics it is easy to f\/ind extensive coverage of the literature, there are many other attempts, that we have tried to present in a way consistent with the increasing depth at which they attack the problem.

The relevance of this kind of research for quantum gravity might be questioned. However, it is a fact that in most approaches to quantum gravity the reconstruction of the continuum macroscopic limit from the underlying microscopic degrees of freedom is a procedure that requires concepts and tools that are common in statistical models\footnote{On this, see the recent application of renormalization group ideas to spinfoam models \cite{bianca}.} and in emergent models. In this sense, we have tried to give a panoramic view also of the dif\/ferent technical tools that might be useful to tackle the problem of the continuum limit of quantum gravity models.

As we have seen, the problem of emerging gravity from some pregeometric nongravitational theory is not yet solved, given that no completely satisfactory model has been worked out.
However, the dif\/f\/iculties encountered do of\/fer also deep insights that are otherwise hardly appreciated in more conventional treatments. Concretely, if gravity is not emergent, but comes from the semiclassical limit of a theory like loop quantum gravity, still the
emergent gravity models might of\/fer insights and techniques to prove such an assertion.

In closing, we would like to address three f\/inal themes.

First of all, it is rather urgent to establish as soon as possible a basis of phenomenological consequences of ideas of emergent gravity, possibly not in the form of Lorentz violation, on which an extensive literature on theoretical models and experimental or observational results exists. A~top-down approach is philosophically satisfactory, but cannot exist without confrontation with experimental evidences from which bottom-up strategies can be elaborated.

{ There is a rather large body of work on Quantum Gravity phenomenology, that we have already mentioned. Much of this work is associated to the possibility that the symmetries that we
observe are just low energy concepts, and they might work dif\/ferently, or being broken, at high energies. Lorentz invariance has of course a special status, in this respect. However, there is much more to do. As we have seen, it is conceivable that dif\/feomorphism invariance is a low
energy symmetry too, as we have discussed in dif\/ferent places. However, being a local gauge symmetry (which enters heavily in the determination of the dynamics in theories like GR), the recovery of a dif\/feomorphism invariant regime requires that the extra degrees of freedom decouple at low energy, or that the nondynamical background structures become undetectable.
The investigation of the potential phenomenology associated to dif\/feo-breaking is a relatively unexplored territory (see \cite{Donoghue}), but it represents an important opportunity to understand how fundamental dif\/feomorphism invariance can be considered, and, in the case in which it turned out to be an approximate symmetry, what is the energy scale that governs the breaking and what are the associated physical phenomena. As in the case of LIV, this will require, most signif\/icantly, the explanation of the (dynamical) origin of the separation of scales between the high energy noninvariant phase and the low energy invariant one, to avoid a naturalness problem. Indeed, in \cite{Nordstrom} this scale turns out to be the Planck scale, but more general situations might be conceived.}

Another possibility is represented by nonlocality, which is a feature that has been conjectured long ago to be underlying certain models in which spacetime and hence locality emerges only on large scales. This is corroborated by many other approaches in which the notion of spacetime arises either as a structure for the propagation of collective modes as phonons in BEC \cite{wroclaw}, or because it arises as a statistical concept as in random models \cite{causets,GFT1, qgraph1}, or because the very topology (in the sense of connectivity) of spacetime looks dif\/ferent at different scales \cite{disorderedlocality1,disorderedlocality2,Sorkinloc}. See also \cite{Giddings:2006sj, Giddings:2006be,Giddings:2004ud} for a discussion of the role of nonlocality in addressing some of the puzzles of semiclassical gravity.

Of course, this is only one possibility among the various possible deviations from the rules of QFT. In general, in pregeometric models, but also in spinfoams and group f\/ield theories, there is one common theme that is the way in which local ef\/fective quantum f\/ield theories that describe the theory in the infrared limit are recovered. This is by no means an easy task, and represents the second theme that is underlying the discussion of emergent models.

Notice also that the fact that, of the various hierarchy problems that plague our ef\/fective f\/ield theories, the cosmological constant problem seems to be less severe in an emergent gravity setting (where the computation of the cosmological constant term is done in term of pregeometrical degrees of freedom) strongly suggests to further investigate the transition between the ef\/fective f\/ield theory regime and the more fundamental layer.

In pursuing this goal, power counting arguments might be of little help. For instance, the argument that Einstein--Hilbert action terms will be dominating the IR action is not guaranteed to be valid, given that it is not automatic that the models will contain such operators in their theory space\footnote{I would like to thank S.~Carlip for having raised this point.}.
For this reasons, models with preferred time, or with certain forms of Lorentz violation might be less compelling than models which, being designed to contain in some form the Regge action for discretized gravity, might encounter more success.

Of course, this is not still a proof that models like group f\/ield theories will deliver the promised result, given that there is the hard problem of the def\/inition of the continuum limit to be solved.
In addition, in pregeometric scenario the very concept of the renormalization group which gives theoretical support to the idea of ef\/fective f\/ield theories has to be reconsidered, in light of the absence of structures with which we can classify modes in terms of IR and UV.
Again, for this purposes, the tools and ideas (especially about the generation of the dynamical scales) that have been developed in emergent gravity  models might be valuable.

The third important theme is the continuous development of tools from areas like condensed matter and statistical systems, that will be key ingredients for many (non emergent) quantum gravity models in the continuum macroscopic limit\footnote{On this, see the preliminary results of
\cite{Oriti2010} and \cite{Critical}.}.

In this sense, the recent use of cold atoms in optical lattices \cite{optical1,optical2} in order to reproduce ef\/fects of quantum f\/ields in extreme conf\/igurations does represent a promising new direction. It has been shown that,
in these systems (which are close to the ones underlying, for the constructive principles, quantum graphity models in certain regimes), many interesting phenomena do occur: the appearance of Dirac operators, abelian and nonabelian artif\/icial f\/ields as well as the possibility of considering phenomena proper of quantum f\/ield theories in extremely intense external f\/ield \cite{szpak2, szpak1} or in which mean f\/ield techniques break down \cite{HFB}. These might provide a valuable set of hints for the identif\/ication of specif\/ic ef\/fects to look for and for the some potential dif\/f\/iculties that we might encounter when abandoning the safe harbor of ef\/fective f\/ield theories to enter the dangerous waters of models in which the notion of smooth background structures might have to be revised.

\subsection*{Acknowledgements}
I would like to thank A.~Campoleoni and F.~Caravelli for discussions, and S.~Finazzi, F.~Girelli,  S.~Liberati and D.~Oriti for the intense collaboration on the subject. I would also like to thank T.~Gherghetta, S.~Hossenfelder, H.~Lin and an anonymous referee for comments, constructive criticism and concrete suggestions for improvements.

\LastPageEnding

\end{document}